\documentclass[acmsmall]{acmart} 
\usepackage{soul}
\usepackage{multirow}

\setcopyright{acmlicensed}
\copyrightyear{2025}
\acmYear{2025}
\acmDOI{XXXXXXX.XXXXXXX}
\acmConference[CSCW '25]{Make sure to enter the correct
  conference title from your rights confirmation email}{XXX}{Bergen, Norway}
\acmISBN{978-1-4503-XXXX-X/2018/06}



\begin{document}

\title[Unpacking Power Among Developers, Designers, and End-Users in FLOSS Usability]{``Ohhh, He's the Boss!'': Unpacking Power Dynamics Among Developers, Designers, and End-Users in FLOSS Usability}

\author{Jazlyn Hellman}
\email{jazlyn.hellman@mail.mcgill.ca}
\affiliation{%
 \institution{McGill University}
 \city{Montreal}
 \state{Quebec}
 \country{Canada}}

\author{Itai Epstein}
\email{itai.epstein@mail.mcgill.ca}
\affiliation{%
 \institution{McGill University}
 \city{Montreal}
 \state{Quebec}
 \country{Canada}}

\author{Jinghui Cheng}
\email{jinghui.cheng@polymtl.ca}
\affiliation{%
 \institution{Polytechnique Montreal}
 \city{Montreal}
 \state{Quebec}
 \country{Canada}}

\author{Jin L.C. Guo}
\email{jguo@cs.mcgill.ca}
\affiliation{%
 \institution{McGill University}
 \city{Montreal}
 \state{Quebec}
 \country{Canada}}

\renewcommand{\shortauthors}{Hellman et al.}

\begin{abstract}
    Addressing usability in free, libre, and open-source software (FLOSS) is a challenging issue, particularly due to a long-existing ``by developer, for developer'' mentality. Engaging designers and end-users to work with developers can help improve its usability, but unequal power dynamics among those stakeholder roles must be mitigated. To explore how the power of different FLOSS stakeholders manifests and can be mediated during collaboration, we conducted eight design workshops with different combinations of key FLOSS stakeholders (i.e., developers, designers, and end-users). Leveraging existing theories on Dimensions of Power, we revealed how participants navigate existing role-based power structures through resource utilization, knowledge gap management, and experience referencing. We also observed that participants exhibited diverse behaviors confirming and challenging the status quo of FLOSS usability. Overall, our results contribute to a comprehensive understanding of the power dynamics among FLOSS stakeholders, providing valuable insights into ways to balance their power to improve FLOSS usability. Our work also serves as an exemplar of using design workshops as a research method to study power dynamics during collaboration that are usually hidden in the field.
\end{abstract}

\maketitle

\section{Introduction}

As efforts worldwide are calling for digital inclusion and software accountability, free, libre, and open-source software (FLOSS)~\cite{feller_framework_2000, feller_perspectives_2005} offers much potential for its values of collaboration, freedom of use, and transparency. This has led to an increase in FLOSS usage and popularity in recent years~\cite{undp_2023}. 
Addressing the usability of FLOSS, however, remains a complex and notably formidable challenge~\cite{cheng_how_2018, hellman_facilitating_2021,andreasen_usability_2006,nichols_usability_2003}. Usability is defined as the ease, error-prevention, efficiency, and pleasantness for an end-user when interacting with software~\cite{cheng_how_2018, nielsen_10_2020}. While many factors contribute to the challenge of FLOSS usability, such as the asynchronous development model and the overwhelming amount and variety of user feedback, a prominent and persistent factor for most FLOSS projects is the long-existing developer-led and development-driven mindset within their communities~\cite{wang_how_2020, hellman_facilitating_2021}.

FLOSS is often embedded with a ``by developer, for developer'' mentality and a culture of prioritizing functionality over usability~\cite{feller_perspectives_2005, wang_how_2020}. While public-facing issue trackers (e.g., GitHub Issues~\cite{cheng_how_2018}) can be used for monitoring usability interests, developers struggle to understand long and convoluted usability discussions~\cite{wang_argulens_2020}. At the same time, designers and end-users often consider issue trackers as obscure and challenging to engage with~\cite{wang_how_2020}. In such contexts, except for a small number of users who are tech-savvy and already have a tight connection with the developers, a vast majority of designers and end-users struggle to influence the FLOSS project. This status quo is often implicit and unbendable, precisely because the roles that experience usability issues and/or have relevant expertise (i.e., end-users and designers) are those with less power to make a difference in FLOSS usability; therefore, they experience difficulties breaking into the FLOSS communities.

To turn the tide to allow active end-user and designer participation in FLOSS, we must first understand their power dynamics with the developers in a collaborative context and consider ways to achieve a more inclusive and democratic process. Here, we interpret power as agencies and ``\textit{a capacity for action}''~\cite{haugaard_four_2021}, building on existing theories of the Dimensions of Power~\cite{lukes1974power,foucault1982subject, Hardy_LeibaOSullivan_1998,haugaard_four_2021}.
This understanding of power provides us with a valuable instrument to examine and mediate the perception and behavior of different roles in FLOSS during collaboration. 

Ideally, we would examine how developers, designers, and end-users collaborate in existing real-world FLOSS development environments. However, end-users and designers are largely invisible in most of these environments~\cite{Hellman_Chen_Uddin_Cheng_Guo_2022,hellman_facilitating_2021, nichols_usability_2003}. As a result, the complex power dynamics of FLOSS stakeholders during collaboration cannot be readily observed or analyzed. To address this issue, and to reveal how FLOSS stakeholders with different roles may navigate through such power dynamics during collaboration, we conceived a design workshop study~\cite{Rosner2016} that brings various FLOSS stakeholder roles together in a collaborative design process.

In particular, we conducted eight design workshops with FLOSS developers, designers, and end-users; each consisted of two or three participants who possessed different roles in FLOSS, totaling 19 participants. Each design workshop was comprised of a moderated focus group and an unmoderated design activity. The focus group sought to precipitate the status quo of FLOSS usability from the participants' perspectives. The design activity had participants work together to create a design pitch for a tool that facilitates the inclusion of diverse stakeholders when addressing usability issues in FLOSS. This design task was deliberately chosen to further tease out participants' reflections and demonstrations of role-based power dynamics during collaboration. 
Further, without moderation by the researchers, participants organically formed the collaborative design setting, such as their decision-making structures and task delegation, which can also influence the participants' behaviors and power dynamics. Thus, the design workshops in our study are used both as a ``field site,'' to reveal the collaborative practices and dynamics of FLOSS stakeholders, and as a ``research instrument,'' to understand how participation and involvement can shape the power dynamics of FLOSS stakeholders~\cite{Rosner2016}.

We performed a qualitative analysis of the design workshop, as well as the resulting design pitches, with a particular focus on role-based power dynamics. 
Our results show that, on the one hand, typical power dynamics in FLOSS (i.e., developers have power over other roles) were replicated during the collaborative design workshops. 
Participants often relied on their tacit understanding of the status quo related to stakeholder roles to inform their design activities. On the other hand, existing power dynamics were also frequently challenged by participants with less role-based power (i.e., designers and end-users) and the status quo de-structured in the participants' vision of the future (through the design pitch). In particular, discursive thought frequently led to changes to the status quo when participants demonstrated empathy, self-awareness of assumptions, or articulated logical reasoning. 

Overall, our work contributes to the understanding of how the existing power dynamics among different FLOSS stakeholder roles might be reinforced and challenged during collaboration to address FLOSS usability. We highlighted ways to overcome these established norms across different power dimensions to build a more democratic process for improving FLOSS usability. Our work also serves as an exemplar for using design workshops as a research method~\cite{Rosner2016} for examining and mediating the power dynamics in collaboration that are usually hidden in the field.

\section{Background and Related Work on FLOSS Usability}
\label{sec:relwork}
In this section, we discuss the collaborative nature of FLOSS and important findings from previous work regarding the tension and challenges concerning FLOSS. We also describe how usability remains an unresolved issue for many FLOSS projects as well as the proposed effort from previous work on improving FLOSS usability.

\subsection{Collaboration in the Context of FLOSS}

FLOSS refers to the software copyright ensuring that \textit{any} person is freely licensed to use, modify, and redistribute the software, sometimes under a certain condition. Since its inception, FLOSS has become a sociotechnical and even political phenomenon, valued by communities who embrace the freedom of exchange~\cite{feller_perspectives_2005}. FLOSS communities are often decentralized and democratic, featuring volunteer-based participation. Historically, projects have followed a loose hierarchy, often called the ``onion model,'' where the repository owner and core developers lead the project (as the center of the onion) while other stakeholders assume more peripheral contributing roles. This model has also represented the trajectory of newcomers to a project where they join through social arenas (such as mailing lists and issue trackers) and become core contributors as they gain knowledge and experience~\cite{crowston2005social}. However, recent research on the social and organizational structures of FLOSS challenges this definition, finding that not all contributors get involved through peripheral activities to become core contributors~\cite {Jergensen2011, Cheng2019-roles}.  

Collaborative activities have a significant impact on a FLOSS project's success. Activities such as social coding, communication, and synchronous development improve productivity and allow for informed decision-making~\cite{dabbish_social_2012, mcdonald_performance_2013, hayashi_why_2013, xuan_building_2014}. However, developers are more likely to collaborate and include contributions from individuals who already demonstrated high-quality work~\cite{tsay_lets_2014}. Additionally, newcomers often have mismatched expectations and strategies with existing contributors~\cite{SteinmacherCGR15,santos2022}. 
Research into what motivates contributors to participate reports both intrinsic motivations, such as altruism, personal needs, and improving programming skills, as well as extrinsic motivations through community identification, career advancement, financial incentives, and investing in the future~\cite{alexander_hars_working_2002, Lakhani_Wolf_2003, von_Krogh_Haefliger_Spaeth_Wallin_2012}. More recent work has found an increasing presence of intrinsic motivations -- altruism grows as contributors gain more experience~\cite{gerosa2021}.

As the number of users and contributors grows and the FLOSS projects evolve, the nature of the FLOSS collaboration changes. More effort from the project team is devoted to maintaining the projects, including effort on user support, establishing trust among contributors, managing the relationship with other projects in a board ecosystem, etc.~\cite{geiger_2021}  During this process, tension and conflict inevitably arise concerning multiple aspects of the FLOSS project and community~\cite{filippova_2015}; these tensions may significantly burden the project maintainers (e.g., the disagreement on software bugs between users and maintainers~\cite{geiger_2021}), while others might lead to positive impacts on the projects, even playing a vital role in ensuring a project's sustainability as they scale. Recursive reflections and resolution of conflicts can result in changes in stakeholders' values~\cite{jamieson_2022}, adjusting technological foundations, and even social foundations of a FLOSS~\cite{filippova_2015}. 

While abundant previous work has studied the dynamics and challenges of FLOSS collaboration, especially from the CSCW and CHI communities, most attention is put on the contributors' and maintainers' perspectives. Such studies often adopted methods such as interviews~\cite{filippova_2015, geiger_2021}, analysis of discussions on FLOSS issues~\cite{GilmerBSCCG23,Hellman_Chen_Uddin_Cheng_Guo_2022}, or a combination of both~\cite{jamieson_2022}. Among those studies, the voice of the designers and users is barely heard, with a few exceptions (e.g., the work by ~\citet{jamieson_2022} and \citet{JahnE0BNMW24}). Moreover, the role of users is mostly framed as the consumers of the technology who might help the project by reporting bugs and engaging in usability testing, rather than part of the collaborator team~\cite{SaneiC24, Iivari09}.

\subsection{FLOSS Usability}

Usability significantly impacts the acceptability and sustainability of the software. Despite being a topic of interest of FLOSS for decades~\cite{nichols_usability_2003, feller_perspectives_2005, cheng_how_2018, rajanen_power_2015}, the usability of FLOSS still falls short. FLOSS has garnered a reputation as being created `by developer, for developer,' sometimes stemming from a mentality of `elitism' and pride over creating hard-to-learn tools negatively impacting usability~\cite{nichols_usability_2003}. In other cases, developers are motivated to address poor usability affecting the users, but they have to juggle with other priorities. In the end, developers can only consider the needs of the core users~\cite{geiger_2021} or resolve usability based on ``common sense'' due to their limited skills and resources~\cite{terry_perceptions_2010, nichols_usability_2003,andreasen_usability_2006}.  
Moreover, in the current FLOSS development context, the discussion on usability typically take place in an ad-hoc manner within a repository's issue-tracking system after the software is released; these discussion threads are mostly text-based and lengthy, containing over-generalized assumptions and focus on developers' needs~\cite{feller_framework_2000, cheng_how_2018, wang_argulens_2020, terry_perceptions_2010}.

User feedback and designer participation are tenants of achieving good usability~\cite{nielsen_usability_1994}. 
However, existing research on usability and participation in FLOSS demonstrates that users and designers generally have very limited decision-making power in the development cycle~\cite{rajanen_power_2015,bach2010, nichols_usability_2003}. The specifics of the decision-making process depend on the size and age of a project. For example, \citet{rajanen_power_2015} found that, in some projects, designers struggled to gain access to decision-making channels due to non-transparent decision-making mechanisms, influencing their degree of impact on the project. While previous work has discussed the practices and challenges of the FLOSS developers and/or designers concerning usability~\cite{nichols_usability_2003, wang_how_2020, feller_perspectives_2005, terry_perceptions_2010, raza_improvement_2010, bach_designers_2009, rajanen_introducing_2012}, only a few investigations have been conducted from `non-core' or `non-technical' end-user perspectives. \citet{hellman_facilitating_2021} found that designers need standardized communication methods, while end-users need inclusive, transparent, and easy-to-learn solutions to be able to participate in the FLOSS development process. User forums, although popular for end-user participation, render end-users with low confidence and decreased power for critical decision-making~\cite{Iivari09, Hellman_Chen_Uddin_Cheng_Guo_2022}. 

We must acknowledge that opening the floor for more inclusive participation from designers and non-technical end-users is not straightforward and requires more nuanced solutions. As revealed in previous work, developers of FLOSS are already struggling with numerous challenges, especially the tremendous burden of handling user requests when the project scales~\cite{geiger_2021}. In this context, it is even more crucial to consider those designers and end-users not as stakeholders who pose challenges when communities grow in popularity and diversify~\cite{wang_how_2020}, but as collaborators to solve issues concerning usability and sustain the thriving community~\cite{Hellman_Chen_Uddin_Cheng_Guo_2022}. FLOSS practitioners need effective means to shift  their mindset and processes to enable such participation and collaboration. Research has demonstrated that power asymmetry between stakeholders in design workshops~\cite{wubishet2013participation} and user forums~\cite{Iivari09, barcellini_situated_2014, iivari_participatory_2011, 10.1145/2948076.2948090} is a key factor in limited participation. Therefore, to design solutions for improving FLOSS usability and increasing the participation of all three stakeholders, we first need an understanding of how power manifests with those stakeholders and how they interact with each other~\cite{dawood_mapping_2019, raza_improvement_2010, rajanen_introducing_2012}.

\section{Theoretical Framework: Dimensions of Power}
\label{sec:power-theory}
The literature on power is well established and discussed in various domains, such as philosophy, psychology, management, and political science~\cite{lukes1974power, Dahl_1957, Kemp_1984, haugaard_four_2021, Allen_2018, Hardy_LeibaOSullivan_1998}. 
We adopted influential theories of the \textit{Dimensions of Power}~\cite{Hardy_LeibaOSullivan_1998, haugaard_four_2021, Haugaard_2012} as the basis for our analysis. Here, \textit{power} is a multifaceted concept and is understood as \textit{``energy, or a capacity for action, which is fundamental to agency. When an agent makes a difference in the social world, they have power''}~\cite{haugaard_four_2021}. Therefore, power is not inherently an oppressive, dominating force that must be eliminated; instead, it is demonstrated through dimensions that make up interactions between actors. 

The theories of \textit{Dimensions of Power} generally extend Luke's theory of power dimensions~\cite{lukes1974power} with Foucault's view of power as an invisible and internalized force within the social fabric~\cite{foucault1982subject} to include four dimensions, as we will discuss below. Modern interpretations of such theories are inspired by feminist theories of empowerment and have been utilized in numerous fields as a way to understand power~\cite{Allen_2018, Bratteteig_Wagner_2016, Farr_2018, haugaard_four_2021, Lobel_2001}. While those dimensions are presented independently, they coexist and interact and intervene with each other constantly. 

\begin{itemize}
    \item \textbf{D-1: Exercises of Power} occur during an interaction between two or more people where someone demonstrates their agency to do something and this action results in a change that would not have happened otherwise. The nature of change depends on whether the exercise of power was authoritative, coercive, violent, or some combination. The type of power exerted depends on how the \textit{resources} are distributed and utilized; such resources can include access to funding and technology, as well as the expertise and experience of those involved. In our study's context, authoritative power would be the most prominent -- one participant exercised their power because they held authority recognized by others.
    
    \item \textbf{D-2: Methods of Structuration} refer to how \textit{social structures} determine whether exercises of power will cause conflict. Exercises of power play out in social structures which are constructed by decisions to include or exclude things (people, norms, and more) from the status quo. This process, called \textit{structuration}, is continuous wherein people maintain, challenge, and/or rebuild these structures. There are two contrasting processes of structuration: (1) \textit{confirm-structuration} when an actor accepts and agrees to continue with the existing social structure, and (2) \textit{de-structuration} when an actor decides not to accept the social structures, therefore challenging the status quo. In de-structuration, there might be conflict if others do not agree with the change.
    
    \item \textbf{D-3: Systems of Thought} capture the process in which underlying knowledge supporting social structures is reasoned, leading to either acceptance or challenge of the social norms. People rely on two systems of thought to accept or challenge social norms. First, \textit{practical consciousness} encompasses the tacit knowledge that is readily known and crucial to the process of structuration. This is how an intuitive understanding of the ``natural order of things'' is derived. Second, \textit{discursive consciousness} represents what a person is actively or ``discursively'' aware of, instead of basing on intuition. When a person is discursively aware of this world order, they can begin to challenge social norms and transform their discursive knowledge into new social norms and a tacit understanding of the world.
    
    \item \textbf{D-4: The Society} refers to how individuals internalize the systems of power in the broader environment that they live in. It occurs because people are constantly being watched and watching others to see what is socially acceptable. Such an environment necessitates the internalization of social norms, making power impossible to escape.
\end{itemize}

\subsection{An Illustrative Scenario in FLOSS}
To understand this theory in the context of FLOSS, we use the following simplified scenario to demonstrate the different dimensions: \textit{a developer contributes a new feature to a FLOSS project through a pull request (PR), but the project owner rejects this contribution.} D-1 refers to the agency and ability of the developer and the project owner to take action. Here, the developer exercises their power to contribute to the feature, causing the owner to react; the owner consequently exercises their power to reject the feature, causing the repository to remain in its current state. D-2 refers to the social structures that give the developer the right and means to contribute through PR and the owner the right and means to reject it. If the developer accepts the owner's decision, then the social structures are maintained. If the developer chooses not to accept the owner's decision, there will be a conflict, such as a discussion or argument in the PR discussion thread. D-3 refers to the tacit social knowledge the developer and the owner both have, such as the project's social structure and the contribution guidelines. If the developer accepts the owner's authority, then they will comply with the owner's decision. However, the developer can also actively think about and challenge the owner's role and authority. Finally, D-4 refers to the overall society, culture, and structures (unique to FLOSS and otherwise) that both the developer and owner internalize and, in turn, use to inform their behaviors. Some examples of this would be following common etiquette or adhering to contribution guidelines because they know their commits and discussions are publicly visible.

\section{Methods}
\label{sec:methods}
To investigate how to improve FLOSS usability and stakeholder inclusion, we conducted eight design workshops. Each workshop had a combination of different FLOSS stakeholder roles to understand their concerns when collaborating with other stakeholders and contributing to FLOSS. This study was approved by the ethics boards of all researchers' institutions.

\subsection{Participant Recruitment and Grouping}
We recruited three types of FLOSS stakeholders -- developers, designers, and end-users -- through advertisements in three primary channels: the Open Source Design forum\footnote{https://discourse.opensourcedesign.net}, the Sustain Open Source Software forum\footnote{https://discourse.sustainoss.org}, and multiple open-source subreddit communities. The ads contained a pre-study survey that asked participants to describe their experiences contributing to and using FLOSS and to self-identify their FLOSS stakeholder roles (i.e., developer, designer, or end-user). In total, we recruited 19 participants, including six developers, six designers, and seven end-users. There were sixteen male participants, two non-binary participants, and one female participant; seven participants were from North America, six were from Europe, four were from Southeast Asia, and two were from Africa. We used the participants' self-identified stakeholder roles to assign them to one of the eight workshop groups; this was to ensure each workshop had different stakeholder roles presented and all workshops covered a variety of stakeholder role combinations. To promote natural discussions during the workshops, we did not disclose the participants' stakeholder roles to other participants prior to the workshops. Table~\ref{tab:workshop_composition} summarizes the workshops and participants.

\begin{table*}[t]
    \centering
    \small
    \caption{Workshop Participants Overview (DEV=Developer, DES=Designer, EU=End-user)}
    \label{tab:workshop_composition}
    \begin{tabular}{lcccc}
        \toprule
        \textbf{Workshop} & \textbf{FLOSS Role} & \textbf{Occupation} & \textbf{Exp. of FLOSS contribution} \\ \hline
        \multirow{3}{*}{W1} & P1DES & Product Designer & 1-5 years \\ 
        & P1DEV  & FLOSS Developer & 30+ years  \\ 
        & P1EU  & Undergraduate Student & -- \\ \midrule
        \multirow{2}{*}{W2} & P2DES & Independent Interaction Designer & 10-20 years \\ 
        & P2DEV & Engineering Professor & 1-5 \\ \midrule
        \multirow{2}{*}{W3}& P3DEV & Recent CS Graduate & 5-10 years \\ 
        & P3EU & Tech Support & -- \\ \hline
        \multirow{2}{*}{W4} & P4DES & Freelance Designer and Animator & 10-20 years \\
        & P4EU & IT Support & -- \\ \hline
        \multirow{2}{*}{W5}& P5DES & {Product Designer \& Researcher} & 5-10 years \\ 
        & P5DEV & {Recent CS Graduate} & 1-5 years \\ \midrule
        \multirow{2}{*}{W6}& P6DEV & {Software Engineer} & 5-10 years \\
        & P6EU & {Recent Civil Engineering Graduate} & -- \\ \midrule
        \multirow{3}{*}{W7} & P7DES & {UX Designer} & 1-5 \\
        & P7DEV & {Technical Lead} & 10-20 years \\
        & P7EU & {PhD Student} & --  \\ \midrule
        \multirow{3}{*}{W8} & P8DES & {IT Coordinator for Schools} & 1-5 years \\
        & P8EUa & {Fine Arts Painter} & -- \\
        & P8EUb & {Web Developer Student} & -- \\ \bottomrule
    \end{tabular}%
\end{table*}

\subsection{Material}
\label{subsec:method_metrial}
Before the workshop, we provided the participants with (1) a set of three personas and (2) a design objective document. Two of the personas, Dakota the Designer and Enrique the End-user, were adapted from Hellman et al.'s work~\cite{hellman_facilitating_2021}. The third persona, David the Developer, is created in this study based on findings from \citet{wang_how_2020}. The design objective document instructed participants to design a new tool for supporting the inclusion of FLOSS stakeholders when addressing usability issues; as a deliverable, each workshop's participants were asked to create a design pitch document. The personas and the design objective document are available in Appendix \ref{appx:material}.

\subsection{Workshop Structure}
All workshops followed the same organization and structure; each lasted about three hours, took place remotely using Microsoft Teams, and was recorded. Participants who provided prior consent were asked to keep their cameras and microphones on for the duration of the workshop. Participants were compensated \$60 CAD for their participation. Each workshop contained the following two main parts.

\subsubsection{Focus Groups}
After introductions, participants reviewed the personas and the design objective. Then, they were asked to discuss, reflect on, and suggest changes to the personas and discuss the design objective. During this process, participants had the chance to share which persona(s) they identified most with; choosing to disclose this information was the only time when participants' roles (i.e., developer, designer, or end-user) were shared with other participants. Persona reflections prompted the participants to consider and share their own experiences participating in FLOSS about usability issues. Through the design objective discussions, participants also discussed how to achieve the objective and measure success.

\subsubsection{Design Activity}
The design activity had two components. Participants first worked together to decide on collaboration tools. Participants were free to choose any collaboration tools to use. The final tools chosen, as well as their discussion process, were connected with the role-based power dynamics and were considered during data analysis. Afterward, the participants worked towards their design objective while the facilitator observed with the camera and microphone off. Participants were unmoderated and had free reign to design any solution they felt satisfied the design objective.

\subsection{Researchers' Reflections on the Workshop Method}
\label{sec:methods_reflection}
In real-world FLOSS contexts, a full picture of the power dynamics among the developers, designers, and end-users is likely not observable. This obscurity stems from stakeholder roles often being opaque to others, and direct communications among roles possibly not happening, depending on the project. 
For example, in real-world settings, end-users often discuss their issues on user forums that are only monitored by very few developers~\cite{Hellman_Chen_Uddin_Cheng_Guo_2022}. Designers have neither dedicated channels to contribute to open source design for FLOSS projects nor to discuss their thoughts and responses; moreover, designers find the existing issue tracking systems hard to use~\cite{wang_how_2020, SaneiC24}. Therefore, in an effort to make the power dynamics observable, we sought to create a setup that would prompt explicit collaborations regarding usability challenges among diverse stakeholders and, at the same time, feel natural for the participants. 

To achieve this, we carefully considered elements in the study design. The workshop format brings diverse stakeholders to a (remote) face-to-face collaboration setting, which does not often occur in the real world, allowing for implicit assumptions and existing beliefs to become explicit, and thus observable, discussions. The design objective itself (i.e., creating a tool to support the inclusion of FLOSS stakeholders to address usability) was deliberately structured to be familiar to the stakeholders and simultaneously disruptive, aimed at triggering participants' reflection and discussion on the status quo. 
As such, using the vocabularies of \citet{Rosner2016}, we used the workshop methods both as a ``field site'' to reveal possible dynamics likely to happen in the real world and a ``research instrument'' to explore how direct participation might shape these dynamics.

At the same time, we made several study design decisions to minimize the influence of the research setting on the participants' natural interactions among themselves. First, to prevent participants from forming presumptions based on roles, at no point did we disclose the participants' self-identified FLOSS stakeholder roles to the others in the workshop. Instead, our focus group prior to the design activities allowed participants to reflect and voluntarily share the persona(s) they identified with and use this as a jumping-off point for further discussions. Second, to mitigate potential biases that our personas could introduce, we intentionally prompted the participants to comment on the accuracy of the personas based on their own experiences. This also allowed us to observe how participants would bring their backgrounds and understanding of the status quo to the table. Finally, the researcher muted herself during the design activity, turned off her camera, and did not moderate in any way. Together, these decisions allowed us to observe collaboration patterns among stakeholders and their power dynamics as they became organically explicit, despite our method of ``fabricating'' a FLOSS collaboration environment. 

Our workshop method is not without any limitations. Although we minimized our influence on the design activities, the participants were still aware of the presence of the researcher, which can lead to continued internalized social norms. Therefore, the fourth dimension of power (i.e., how individuals internalize the systems of power in the broader environment of FLOSS) cannot be accurately observed through the workshop in our research site. Despite this limitation, the workshop is a unique setup and effective mechanism for studying most collaboration-based power dynamics of FLOSS stakeholders regarding usability challenges. 

\subsection{Data Analysis}
Each workshop yielded an approximately three-hour-long video recording, the associated verbatim transcript, and the proposed design pitch. We performed a qualitative analysis of the following three types of data.

\subsubsection{Analysis of the Design Pitches}
To analyze the pitches, we focused on the text and visual components submitted by the participants at the end of the study. Two researchers independently coded the pitches for key feature categories. The researchers then discussed, merged, and refined the identified categories, which focused on the overall theme, scope, and key features of the pitches.

\subsubsection{Analysis of the Focus Group}
We performed a thematic analysis~\cite{Vaismoradi2013} on the focus group transcripts to identify themes on the status quo of FLOSS usability. First, three researchers independently completed a round of inductive coding on transcripts from different focus groups. They then discussed, merged their analysis results, and identified common themes. These themes captured participants' perceptions of the current challenges and barriers to FLOSS usability.

\subsubsection{Analysis of the Design Activity}
This analysis aims to surface the underlying mechanisms of participant behaviors that demonstrate the presence and utilization of power based on their roles during the design process. We first defined the unit of the analysis as a section of the design activity in which participants performed a series of actions to achieve a distinct \textit{goal}, referred to as an \textit{episode}. Four researchers coded the workshop videos and identified a total of 599 episodes; the workshops had an average of 75 episodes (minimum of 46, maximum of 207). We then categorized each episode into one of the two groups: \textit{working episodes}, where participants actively created, designed, wrote, and/or implemented the final design pitch; and \textit{discussion episodes}, where participants focused on verbal exchanges of ideas to achieve the episode's goal. During this process, 208 working episodes and 391 discussion episodes were identified. Due to the nature of remote collaboration, there was insufficient data to investigate power dynamics during \textit{working episodes} (e.g., limited body language, not sharing their screen as they work, and a lack of dialogue). Therefore, power dynamics were analyzed only in the \textit{discussion episodes}.  

Specifically, we conducted a reflexive thematic analysis that organized the codes along the first three dimensions of power presented in Section~\ref{sec:power-theory}~\cite{lukes1974power, foucault1982subject, Hardy_LeibaOSullivan_1998, haugaard_four_2021} to understand how each dimension manifested in the discussion episodes. The fourth dimension is not considered in our analysis considering the nature of the workshop. As discussed in Section~\ref{sec:methods_reflection}, a researcher was obliged to conduct and observe the participants during the study, so that the participants always knew they are being observed. As a result, any observation on the fourth dimension would be influenced by the researchers, confounding the results.

A single episode could be coded to have any combination and amount of dimensions (D-1, D-2, D-3). First, one author independently coded two workshops and discussed the analysis with the other researchers. Then, three authors derived a structure for the three power dimensions in our design activity and created the code book. Finally, one of the three authors who created the code book and another author divided up the task of coding the remaining workshops and refining the code book as needed.

\vspace{10pt}
The analysis results are presented in the next sections. To provide a big picture for contextualizing the participants' power dynamics during collaboration, we first summarize the design pitches created during the workshops (Section~\ref{sec:design-pitches}) and report the participants' reflections on the status quo of FLOSS usability that arose during the focus group portion of the workshop (Section~\ref{sec:status-quo}). We then closely examine the role-based power dynamics during collaboration and describe how collaboration influenced the dynamics in turn. This is achieved through reporting our thematic analysis results on the observed power dynamics across the three power dimensions (Sections~\ref{sec:d1-results}, \ref{sec:d2}, and \ref{sec:d3}, respectively).

\section{Results: Design Pitches Created in the Workshops}
\label{sec:design-pitches}

Table~\ref{tab:design-pitches} summarizes the characteristics and key features included in the design pitches. For collaborating, the groups used various tools, including Miro\footnote{https://miro.com/}~[\textit{W1, W2, W7}], Google Doc~[\textit{W3, W4, W7, W8}], local HTML editor (with screen sharing)~[\textit{W3}], Figma\footnote{https://www.figma.com/}~[\textit{W4}], Riseup Pad\footnote{https://pad.riseup.net/}~[\textit{W5}], Excalidraw\footnote{https://excalidraw.com/}~[\textit{W6}], LibreOffice\footnote{https://www.libreoffice.org/}~[\textit{W7}], and paper and pencil (taking pictures and uploading to document)~[\textit{W7}].
Below, we briefly describe each design pitch created by the participants.

\begin{table*}[t]
\caption{Summary of design pitch key features resulted from each workshop. Features were marked as being present in a pitch \textit{only} if it was visually present in the final submission (i.e., if the feature was discussed but not included in the final pitch, it was not marked).}
\label{tab:design-pitches}
\resizebox{\textwidth}{!}{%
\begin{tabular}{llcccccccc}
\toprule
\multicolumn{2}{l}{\textbf{Feature Categories}} & \textbf{W1} & \textbf{W2} & \textbf{W3} & \textbf{W4} & \textbf{W5} & \textbf{W6} & \textbf{W7} & \textbf{W8} \\ \midrule
Main focus & Facilitating Community Building & \multicolumn{1}{c}{X} &  &  & \multicolumn{1}{c}{X} & \multicolumn{1}{c}{} &  & \multicolumn{1}{c}{} &  \\
 & Facilitating Communication &  & \multicolumn{1}{c}{X} & \multicolumn{1}{c}{X} &  & X & \multicolumn{1}{c}{X} & X & \multicolumn{1}{c}{X} \\ \midrule
Scope & Central platform hosting multiple FLOSS projects & \multicolumn{1}{c}{X} &  &  & \multicolumn{1}{c}{X} &  &  &  &  \\
 & Tool integration into one single project &  & \multicolumn{1}{c}{X} & \multicolumn{1}{c}{X} &  & X & \multicolumn{1}{c}{X} & X & \multicolumn{1}{c}{X} \\ \midrule
Communication & Synchronous &  &  &  &  & \multicolumn{1}{c}{} & \multicolumn{1}{c}{X} & X &  \\
 & Asynchronous &  & \multicolumn{1}{c}{X} & \multicolumn{1}{c}{X} &  & X & \multicolumn{1}{c}{X} & \multicolumn{1}{c}{} &  \\
 & Familiar modes of communication for end-user (e.g., Slack, SMS, video) &  & \multicolumn{1}{c}{X} & \multicolumn{1}{c}{X} &  & X & \multicolumn{1}{c}{X} & X & \multicolumn{1}{c}{X} \\ \midrule
Features & Accessibility (e.g., closed captioning, accessibility support groups) &  &  &  & \multicolumn{1}{c}{} & \multicolumn{1}{c}{} &  & X & \multicolumn{1}{c}{X} \\
 & Translation features &  &  &  & \multicolumn{1}{c}{X} & & \multicolumn{1}{c}{} & \multicolumn{1}{c}{X} & \\
 & Telemetry collection &  &  &  &  & X & \multicolumn{1}{c}{X} & \multicolumn{1}{c}{} & \multicolumn{1}{c}{X} \\
 & AI facilitated solution (e.g., chatbot, Discord bots, etc.) &  &  &  &  & X & \multicolumn{1}{c}{X} & X & \multicolumn{1}{c}{X} \\
 & Feedback collection for usability issues &  & \multicolumn{1}{c}{X} & \multicolumn{1}{c}{X} &  & \multicolumn{1}{c}{} &  & X & \multicolumn{1}{c}{X} \\
 & Multi-media collection in user feedback forms (e.g., audio, screen recordings) &  & \multicolumn{1}{c}{X} &  &  & \multicolumn{1}{c}{} &  & X &  \\
 & GitHub integration & \multicolumn{1}{c}{X} & \multicolumn{1}{c}{X} &  &  & X & \multicolumn{1}{c}{X} & X &  \\
 & Life cycle management (e.g., Kanban, roadmaps, etc.) & \multicolumn{1}{c}{X} &  &  & \multicolumn{1}{c}{X} & X &  & \multicolumn{1}{c}{} &  \\
 & Inclusive documentation, resources, \& education & \multicolumn{1}{c}{X} &  &  & \multicolumn{1}{c}{X} & \multicolumn{1}{c}{} &  & X & \multicolumn{1}{c}{X} \\ \bottomrule
\end{tabular}%
 }
\end{table*}

All the workshops created design pitches that either facilitate community building or communication. \textit{W1} and \textit{W4} created \textbf{central hubs across FLOSS projects to promote community building}. They included features for attracting potential collaborators, connecting people with existing projects, making resources (e.g., documentation and contribution practices) more accessible, and abstracting away from GitHub to encourage non-developer community participation. \textit{W2, W3, W7,} and \textit{W8} focused their solution on \textbf{facilitating feedback-related communications within a FLOSS project}, specifically bug reports, usability issues, and feature requests. These pitches proposed tools that aim to lower participation barriers for non-technical end-users and facilitate communication among stakeholders. They included features such as multi-media feedback inputs, synchronous communication channels, chat-bots for feedback collection, accessibility support, structured chains of communication (e.g., end-users to designers to developers), and GitHub integration. \textit{W5} and \textit{W6} also designed for communication for individual projects, but focused on \textbf{streamlining the information flow among stakeholder roles}. \textit{W5} proposed a GitHub plug-in for notifying other stakeholders when to join a conversation while \textit{W6} proposed a formalized feedback loop to streamline usability information transfer between any two pairs of stakeholder roles. Both pitches included features that allowed for communication external from GitHub (e.g., through Slack, email, etc.) and incorporated the use of bots.

\section{Results: Participants' Reflections on the Status Quo of FLOSS Usability}
\label{sec:status-quo}

During the focus group portion of the workshop, participants reflected on their experiences and challenges when addressing usability in FLOSS. We identified four main themes from their discussions that we describe below. 

\textbf{Communication Barriers.} Participants all felt that communication-related factors were a common issue hindering usability and participation. These factors included challenges surrounding miscommunication, jargon and technical terminology, and communication channels that require technical knowledge. End-user participants experienced hostile communication styles and expressed an overall lack of confidence due to limited knowledge of terminology, norms, and processes. For example, P1EU said that they ``\textit{[don't] really know how to ask a question ... [or] even, what's going wrong. And that kind of generates a lot of confusion.}'' Developer participants experienced this, too, and discussed how FLOSS communication relies on technical knowledge that might overwhelm many end-users (e.g., issue tracking systems, detailed issue templates, and Internet Relay Chat (IRC) systems). Developer participants also expressed the time-consuming and challenging nature of interpreting user feedback. Some designer participants believed that communication challenges arise because developers lack the skills to engage with user feedback. Participants agreed that communication channels needed to become more inclusive. P4DES corroborated these sentiments, saying that \textit{``technical friction should be removed for users that want to contribute.''}
 
\textbf{Obscure Documentation.} Similarly, participants agreed that there was a need for more inclusive documentation so non-technical end-users could become involved in FLOSS. Some participants said that the status quo lacks specific onboarding documentation and that, typically, documentation is solely for developer usage. For example, P1DES noted a situation where they reached out to a project for usability documentation, to which a maintainer responded by saying: ``\textit{most of [that project's] current docs are majorly focused just on developers.}'' P1DEV and P1EU directly echoed this sentiment. This inhibits designers' and end-users' opportunities to contribute to FLOSS and improve a project's usability.

\textbf{Development-Centric Culture.} All participants had encountered a community that treats usability as an ``add-on'' feature, focusing on functionality over usability. For example, P3DEV said that, in their experience, a major problem with FLOSS projects is they tend ``\textit{to fulfill the needs of the developer more than the users}'' and that there is ``\textit{some selfishness}'' involved when developing these features. Participants believed this development-centric culture was largely responsible for users' lack of trust in the participation process, presuming that their input would not be valued, and that this generally caused usability to become an afterthought. However, as P1DES discussed, the traditional \textit{``scratching an itch''}  approach (i.e., choosing what to/not to fix based on developers' own interests) cannot scale well with more users. Moreover, participants discussed how the status quo of FLOSS is not focused on improving barriers to community building, but that it should be. Specifically, designers and other non-developers are typically not included in a project pipeline.

\textbf{Lack of Incentives.} Finally, participants reflected on the role of incentives in FLOSS, identifying that the status quo lacks incentives for developers to prioritize usability and for designers and end-users to contribute to FLOSS projects. Financially, designer and developer participants expressed that donation buttons are insufficient because they are not always properly distributed to contributors. Moreover, despite end-users expressing their desire to participate and help FLOSS, they do not always feel motivated because the software is free. For example, P3EU said they sometimes felt embarrassed asking developers for more because ``\textit{they developed the software for free. So, who am I to complain?}''

The topics discussed by the participants were based on their own experiences in FLOSS, illustrating that the problems in FLOSS usability remain pervasive. Such discussion during the focus group enabled the participants to learn from each other, understand and consider what challenges other stakeholders face, and enter the design activity with a clear understanding of what issues are most important to everyone in the workshop.

\section{Results on D-1: Exercises of Power in the Design Workshops}
\label{sec:d1-results}

From Section~\ref{sec:d1-results} to \ref{sec:d3}, we report the power dynamics identified during the participants' collaboration for the design activity. In our analysis, we interpret the participants' power dynamics by comparing their actions and decisions with the status quo of FLOSS usability reported in Section~\ref{sec:status-quo}. 

The first dimension of power refers to the \textit{exercises of power} that cause any change in the world (see Section~\ref{sec:power-theory}). We identified the following three key processes comprising the first dimension, which allowed participants to define their agency and exercise power to accomplish the episodes' goals. 

\subsection{Resource Utilization}
\label{sec:d1-res-util}
The first process occurred when a participant demonstrated agency to achieve a goal through \textit{resource utilization}. The potential power of a participant was defined by the available resources and their ability to use them. The process of using resources, in turn, directly impacted the design activity and its final outcome. Specifically, we observed three types of resources. 

\textbf{Utilization of workshop artifacts.} Workshop artifacts were documents or materials either provided by the research team, such as the personas (see Section~\ref{subsec:method_metrial}), or created by the participants during the workshop. Participants used these artifacts to inform opinions, negotiate ideas, and support decision-making. For example, in \textit{W1}, P1DES communicated their design decisions to their teammates by using design artifacts created earlier in the design activity, i.e., sticky notes identifying key problems in FLOSS. These artifacts were used as a visual aid to explain and justify the design. As P1DES explained the pitch's key features, they selected the relevant sticky notes and moved them into a new pile, thus exercising their power to make critical decisions on the design pitch. In another example, at the very start of \textit{W8}, P8EUa was the first to speak and hesitantly suggested they revisit their focus group discussions to identify a problem. P8EUa continued to brainstorm and frequently referred back to the personas to draw ideas and support their opinions and suggestions. Increasingly, we observed that they expressed their thoughts and ideas more confidently.

\textbf{Utilization of external resources and documentation.} Participants obtained resources external to the workshop, enabling exercises of power to achieve goals. These external resources were used as aids when communicating ideas, answering questions, providing justifications, etc. For example, in \textit{W4}, P4EU searched online and found a platform with a concept similar to the proposed idea. P4DES considered P4EU -- who found the resource -- the authority on this resource and asked P4EU if it differed from their idea. This resource allowed P4EU to continue contributing to their original idea confidently and guide the team's pitch direction. 
Similarly, in \textit{W3}, P3EU -- whose participation was limited due to P3DEV using a local HTML editor -- independently searched for an external resource for the design pitch while P3DEV worked on their own. P3EU informed P3DEV: \textit{``I've just linked to an external site I saw of best practices for feedback forms, so maybe this will be helpful.''} While no further discussions occurred about these best practices, the link was included in their final pitch.

\textbf{Utilization of skill sets.} Participants leveraged various skills during the design activity to achieve their goals, including skills with respect to collaboration tools (e.g., Miro, Google Doc, etc.), development, and design. When participants had skills in any of these areas, they were more likely to act and exercise their power. When one participant had limited skills compared to another, it led to: (1) restricted or no agency to participate, or (2) their exercise of power in some other way to include their perspective. For example, in \textit{W3}, P3DEV decided to create the UI with a local HTML editor while screen-sharing. P3DEV then had greater power over the pitch than P3EU, who did not have development skills or access to the editor. As such, on only one occasion, P3EU suggested an idea, which was including non-technical roles (i.e., user) as a drop-down option in the feedback form (see Section~\ref{sec:design-pitches}). On the other hand, in \textit{W7}, the three participants struggled to agree on a collaboration tool. Ultimately, to accommodate all skills, the three participants decided to work individually with their preferred tools, allowing each participant to exercise their power to propose and share designs. 

\subsection{Managing Knowledge Gaps}
\label{sec:d1-knowledge-gap}
The second process occurred when one participant was recognized as an authority for the knowledge they possessed that others did not have. Managing knowledge gaps resulted from this unequal distribution of knowledge \textit{resources} (i.e., skill), influencing the participants' perceptions of each other. In turn, this determined the power dynamics between them and their behaviors to achieve the goals. This process transpired when one participant engaged in a conversation; this process was identified when a participant exhibited one of the following behaviors: 

An \textbf{inhibition behavior} represents a process in which the participant was not an authority and they elected not to exercise power to obtain the missing resources from the authority. Inhibition behaviors further limit one's capacity to act. For example, in \textit{W1}, P1EU suggested an idea but then said: \textit{``I'm not a developer, I don't have a clue what I'm saying''} and didn't ask P1DEV for their perspective, exhibiting a limited capacity to act in this context. For another example, in \textit{W8}, P8EUa struggled repeatedly to concretize their chatbot idea for improving stakeholder communication. This was evident in a few instances when they said, \textit{``I don't know where the chatbot comes in,''} or struggled to formulate thoughts on chatbot implementation. Although \textit{W8} did not have a developer participant present, P8EUa did not attempt to improve their capacity to act through gaining information from alternative sources. Although P8EUa initially had the power to participate, the missing knowledge ultimately limited their agency as P8EUa concluded the episode with an undeveloped idea.

A \textbf{motivated growth} represents a process in which the participant was not an authority but obtained the missing knowledge, thereby increasing their potential power in future episodes. For example, in \textit{W2}, P2DEV, who had limited prior knowledge of usability concepts and, consequently, had limited potential to act, counted P2DES as an authority on usability and asked for clarification on usability issues. P2DEV asked P2DES to clarify if \textit{``categoriz[ing] the issue as usability problem or not[, means] we accept this as a thing for us to work on or ... or not, right? Or is it different from that?''} By asking for clarification, P2DEV demonstrated their treatment of P2DES as the authority on usability in addition to exercising their agency to obtain information and make informed decisions on the topic moving forward. As another example, in \textit{W1}, P1DEV suggested ideas related to linking a project's issues and feature requests to the road map. P1EU then asked P1DEV : \textit{``So, is a road map a feature of GitHub? I don't know all of its ...''}. After P1DEV responded about road maps and GitHub, P1EU participated more actively in brainstorming on P1DEV's initial ideas.

A \textbf{tutoring behavior} represents a process in which the participant was an authority on the subject and exercised their power to disseminate knowledge to another participant. This behavior caused the other participants, who accepted them as an authority, to react accordingly. For example, prior to the earlier example episode in \textit{W2}, P2DES and P2DEV were identifying existing problems in FLOSS when P2DES chose to explain the nuance surrounding usability issues and share examples. Here, P2DES was not prompted to explain any concepts but did so spontaneously, indicating their agency to share the usability information as the authority and resulting in P2DES's power over how the information was disseminated. Another example in \textit{W5} discussed a well-known issue in FLOSS: tool adoption. P5DES described their idea as a plugin for existing software rather than a new tool entirely. P5DES then explained to P5DEV that their rationale surrounded the difficulties that arise when getting FLOSS users to switch software for the sake of improved communication. P5DES shared this knowledge with P5DEV unprompted, exercising their power to ideate and guide the design pitch.

\subsection{Referencing Past Experience}
\label{sec:d1-past-exp}
The final process occurred when a participant referenced a past experience about using a \textit{resource}. Participants engaged in this process by explicitly mentioning or describing an experience in the past, signaled through the use of a variation of the phrase \textit{``in the past'', ``I've used [X] before'',} or \textit{``we did [x] in the past''.} Past experiences were utilized in two ways.

First, a \textbf{justification behavior} occurred when participants communicated a past experience to justify a decision made during an episode. For example, in \textit{W5}, P5DES identified lack of accessibility as a problem for the end-user persona when P5DEV asked: \textit{``What does that FLOSS even look like? [A] FLOSS for coordinating volunteers [with special needs] through web searching?''}  To answer the question and justify that such a FLOSS is possible, P5DES provides an example by sharing: \textit{``So last week I was coaching the project which is -- this one! -- This one's: `team organization for spontaneous help,' so spontaneous, like, volunteers.''} As P5DES justified the concept by sharing their past experience, they also utilized external resources to support their argument. P5DEV is convinced, and the pair continue designing. As another example, in \textit{W6}, P6EU received pushback from P6DEV on adding a field in bug reports to support different end-user categories. In response, P6EU shared: \textit{``I have seen people saying, `It said this is in the app description, but it's not.' So if a feature is somewhere hidden or in the menus, users sometimes really don't do proper checks.''} P6EU used this experience to justify their idea, to which P6DEV tentatively agreed.

Second, an \textbf{authority establishment behavior} occurred when participants shared the use of a resource in the past to demonstrate and establish authority, either intentionally or consequentially, by illustrating their credibility and trustworthiness. For example, in \textit{W1}, P1DES suggested an idea for implementing regular synchronous meetings in their pitch, towards which P1EU expressed some doubt. In response, P1DES argued to include it by responding: \textit{``Yeah, [my FLOSS company] does it well,''} and explaining the company's logistics, successfully establishing their credibility and authority on the topic. Similarly, in \textit{W5}, when brainstorming a feature to provide decision-making contexts, P5DES communicated their authority by comparing their own experience with that of a persona: \textit{``Frequently, I've been in the phase that Dakota's been in where I'll be researching [design] options. Then suddenly the development team chose a library without communicating...''} To this, P5DEV responded, \textit{``[My company is] working on something like this. It is used internally. It's a huge context store. ... So I feel like we are thinking of something similar.''} In doing so, P5DEV accepted P5DES's authority and simultaneously established their own credibility.

\section{Results on D-2: Methods of Structuration}
\label{sec:d2}
The second dimension of power concerns \textit{structuration}, the method through which people maintain, challenge, and rebuild social structures (see Section~\ref{sec:power-theory}). In our analysis of the design workshops, we identified key behaviors the participants exhibited for both \textit{confirm-structuration} (i.e., participants continued with the existing FLOSS social structure) and \textit{de-structuration} (i.e., participants challenged the existing social structure). Here, we use the following terminology: participants with \textit{more power} and participants with \textit{less power}; this refers to the stakeholder group the participant belongs to and where in FLOSS's hierarchy of power they traditionally fall. FLOSS developers traditionally have the most power, followed by designers and then end-users~\cite{rajanen_power_2015}. Note that \textit{a participant with more power} does not necessarily mean they exercise their power more frequently. The two structuration processes occur when a participant's behaviors in the design activity confirm or challenge the traditionally established power hierarchies among stakeholder roles. Below, we describe how the behaviors of both processes are demonstrated.

\subsection{Confirming Social Structures}
\label{subsec:d2_confirm_structure}
Confirm-structuration occurred when participants upheld the status quo of FLOSS (see Section~\ref{sec:status-quo}), either during collaboration or reflected in their design ideas. 

\subsubsection{Confirm-structuration during collaboration}
When participants collaborated, two key behaviors confirmed the FLOSS social structures.  

First, sometimes, the overall \textbf{decision-making was driven by the participants with more power} during the design activity. This occurred when the participants with more power ignored or corrected the other participants, independently decided on the collaboration tool, led the tasks during the design activity, made the final decisions about what was included in the pitch, or acted as the main contributor to the pitch. These behaviors were demonstrated by DEV and DES participants, confirming the status quo that more powerful stakeholders remain the primary decision-makers. For example, in \textit{W4}, a confirm-structuration episode occurred when P4EU suggested an idea, then P4DES responded with: \textit{``I think it should be a feature, but it shouldn't be baked in''} without further discussions. P4EU agreed immediately, replicating existing FLOSS structures where the stakeholders with more power, here, the designer, had the final decision-making power. 

Second, the confirm-structuration behavior was also demonstrated by the participants with less power when they \textbf{deferred to a participant with more power} by seeking their approval, accepting their ideas without further discussions, providing unprompted updates, and asking for help from a participant with more power. For instance, in \textit{W3}, P3EU deferred to P3DEV about the design pitch by asking: \textit{``What do you think are the key features of our simple feedback form? Is it easy to use and figure out?''} As P3DEV responded, P3EU accepted the responses and documented the answers in the written pitch. Similarly, in \textit{W5}, P5DES was synthesizing the discussions on the main problem to address and sought confirmation from P5DEV: \textit{``Is that like ... the problem or the meta-problem that we're focusing on? ... Have I understood?''} P5DEV said \textit{``Yeah''} and provided some more insights.  

\subsubsection{Confirm-structuration in design ideas}
Participants also proposed and prioritized features for their design pitch (Section~\ref{sec:design-pitches}), that reflected the existing FLOSS status quo categories.

First, participants proposed ideas that intentionally or unintentionally \textbf{created communication barriers among FLOSS stakeholder roles}, reinforcing norms where developers, designers, and end-users are separated or divided throughout the project lifecycle rather than actively collaborating, causing communication struggles. For example, in \textit{W8}, P8EUa suggested a bug reporting feature for end-users; when asked for clarification, P8EUa elaborated: \textit{``I assume that [for a bug reporting tool,] one side of it would be for people like Enrique, uh people who are less technically savvy. But on the other side of it, I assume would be David and Dakota.''} This feature reflected the current FLOSS social norms that separate software consumers (end-users) and software creators (developers/designers). 

Second, some participants \textbf{prioritized technical feasibility} over usability when ideating features for the design pitch or considered feasibility a major concern in the pitch. Remaining in this feasibility mindset confirm-structured the norm of a \textit{development-centric culture}. For example, when P1EU began to discuss implementing a front-end interface to simplify GitHub repository information, they eventually switched gears and argued against including this idea. They perceived the effort and resources required for implementation as too high, despite not actually needing to implement the design pitch at the end of the workshop. Similarly, in \textit{W7}, after the participants identified the main goals they wanted to address, P7EU said: \textit{``We covered most of the important points and now we should consider feasibility... for developing the tool.''} This demonstrated a mindset that feasibility considerations are critical for moving forward.

Finally, participants \textbf{prioritized developer needs} when ideating the design pitch, also confirming the \textit{developer-centric culture} status quo. Notably, this occurred when design features dictated developers as sole decision-makers and prioritized existing developer-centric norms (e.g., no usability support) or tools (e.g., GitHub, issue tracking, telemetry analysis solutions). For example, in \textit{W5}, P5DES was discussing a mechanism for handling different issue types when they suggested: \textit{``Maybe it's just like a pause and a rewind and a stop button on issues in GitHub [because] they need a context check [and] it's kind of like a plugin for existing tools.''} P5DES then justified their idea as a plug-in centering around GitHub due to switching costs and new tool adoption challenges. While a valid justification, the proposed solution confirms GitHub-centric norms. As another example, in \textit{W3}, P3DEV created a drop-down menu for a person to self-identify as \textit{``developer, user, or tester''}, ignoring P3EU's earlier suggestion to include designers. Later on, P3DEV also informed P3EU: \textit{``I created another drop-down that has [options for] feature request, bug report, and others.''} This is a feature similar to those existing of GitHub and does not consider designers' and end users' needs.

\subsection{Challenging Social Structures}
\label{subsec:d2_de_structure}
The other process, de-structuration, occurred in episodes where participants challenged or changed FLOSS' status quo during collaboration or through design ideas.

\subsubsection{De-structuration during collaboration}
When participants collaborated, they performed the following behaviors challenging the FLOSS social structures.

First, sometimes, \textbf{participants with less power guided the design activity} by acting in ways typically performed by participants with more power, therein changing the dynamics of participation and decision-making typically reflected in FLOSS. This occurred when participants, often end-users, decided the pitch's design direction, served as the main contributors of ideas, guided the strategy for the team, or suggested alternative ideas that challenged or enlightened a participant with power. For example, at the beginning of \textit{W8}, P8EUa commenced the activity by suggesting: \textit{``should we talk about what specific problems we found?''} P8DES and P8EUb agreed and the team immediately continued with identifying problems they would like to focus on. P8EUa demonstrated their agency to guide the initial phases of the design activity -- a task usually performed by the DES or DEV participants. In another example, in \textit{W2}, when the participants finished ideating, P2DES suggested they begin wireframing and invited P2DEV to participate: \textit{``If there is something here [on the Miro board], just jump in and change things.''} Through this process, P2DES's design skills allowed them to guide the activity and de-structure the FLOSS decision-making norms.

Second, a participant with power also de-structured collaboration norms when they \textbf{deferred to a participant with less power}, by accepting ideas proposed by participants with less power or seeking input, approval, or confirmation about a proposed strategy from them. For example, in \textit{W2}, P2DEV began to communicate some updates they made to the design pitch when they paused and said to P2DES: \textit{``I don't know, [there's] probably better ways to do it.''} This indicated that P2DEV wanted input from P2DES on their work and de-structuring the status quo. Similarly, in \textit{W5}, P5DEV shared their thoughts on a potential solution with P5DES and asked P5DES for their opinion on whether to include it in the design pitch. P5DEV concluded with, \textit{``Or maybe you can improve on it...?''} By deferring to P5DES in this way, P5DEV de-structured the collaboration norms in FLOSS that developers are often the final decision makers.

Finally, when collaborating, a participant sometimes \textbf{demonstrated knowledge or skills that were outside of their usual stakeholder roles}. This often occurred when an end-user participant demonstrated developer or designer knowledge, or a developer participant performed design activities (i.e., used wireframing tools). For example, in \textit{W8}, P8EUa indicated that they don't know how feasible implementing a chatbot would be. While P8DES responded by saying: \textit{``that's the fun part of this workshop is that we are free [to brainstorm],''} P8EUb then commented on the feasibility aspect. Here, P8EUb said: \textit{``I also don't know how feasible it would be, but a lot of FLOSSes are based off of modularity,''} briefly explaining what that means and its potential. Although P8EUb acknowledged their limited knowledge, they still exercised their power to share information and help decision-making by assuming authority on more technical matters. Similarly, in \textit{W7}, P7DES asked if anyone was familiar with Figma. P7DEV said they were not, but added: `\textit{``I can do sketching and Photoshop and share the screen if it helps.''} P7DES then suggested working individually and later regrouping. Through this collaboration strategy, both P7DEV and P7EU then exercised their power to create unique designs that were all included in the final pitch.

\subsubsection{De-structuration in design ideas}
Participants also worked to de-structure the FLOSS norms when the design pitch features facilitated stakeholder interaction and integration through the following ways.

First, participants changed or challenged the status quo by \textbf{removing communication barriers} in FLOSS. This occurred when ideas and features in the design pitch de-structured \textit{communication barriers} through synchronous communication features (i.e., video calls) and communication methods that didn't rely on technical knowledge. For example, in \textit{W7}, although each participant performed their design separately, all designs included a user feedback tool that allowed for both asynchronous and synchronous communication. P7DEV and P7DES included \textit{``video chat''} features and P7EU included a \textit{``synchronous audiovisual component''} feature. P7DES further explained: \textit{``[The feature] can serve as output for audio and video calls, which you see on the next screen -- just like [this Microsoft Teams call here] and there is a chat box if you're not able to communicate. So that solves the problem of accessibility.''}  

Second, some participants created features in the design pitch that \textbf{prioritized inclusive documentation} to combat the FLOSS norm of \textit{obscure documentation}, such as including features for easy-to-find documentation, design contribution guidelines, and onboarding documentation for diverse stakeholders (e.g., end-users, designers, technical writers, artists, etc.). This was exemplified in \textit{W1} where P1DES, who had finished their visual designs, communicated the inclusive and easy-to-find documentation features to the other participants: \textit{``The docs can be linked and then, if you are a designer, probably they are going to make a contribution guideline. [...] It's just a click-through. You click, you go to the next thing. `Ohh' -- you see what you need. They probably link to the Figma file or then the docs.'' } P1DEV and P1EU agreed on these designs, which were included in the final design pitch. A similar situation was seen in \textit{W7} when the participants were identifying existing problems to focus on, P7EU said: \textit{``Perhaps we can think about the design of the documentation for the user... -- so that it is not visually scaring someone who is not very knowledgeable.''} This was further discussed by the team and the feature was eventually included in the pitch. 

Finally, participants also changed or challenged the \textit{development-centric culture} of FLOSS by \textbf{prioritizing usability}. This occurred through discussing and including features for user-friendly designs that abstracted away technical components and considered non-developer needs, such as through designer or end-user-centric platforms. For example, in \textit{W6}, P6DEV advocated for less confusing bug reporting processes, specifically so end-users would be more comfortable getting involved. P6DEV envisioned helping users by abstracting the technicalities of reporting different types of bugs, saying: \textit{``I don't think users know what's technical and what's not ... So, we have to design it in such a way that a user can report things without getting confused.''}
\section{Results on D-3: Systems of Thought}
\label{sec:d3}
The third dimension of power concerns the driving forces behind the participants' behaviors and their underlying cognitive processes as they exercise power and affect the status quo (see Section~\ref{sec:power-theory}). We analyzed the episodes for the two types of thought that are used during exercises of power: (1) their practical consciousness (i.e., source of ``meaning'' and subconscious reflected in actions) and (2) their discursive consciousness (i.e., active thought encompassing goals and objectives). 

\subsection{Practical Consciousness}
\label{sec:d3_practical_conscious}
Specifically, practical consciousness ``refers to the vast tacit knowledge which actors use to inform their structuration practices''~\cite{Haugaard_2012}; it is immediately known knowledge of how the world works, is often ``taken for granted,'' and allows the natural order to carry on (i.e., confirm-structuration). 
We identified two ways the participants demonstrated practical consciousness as they worked through the activities. 

First, practical consciousness was demonstrated \textbf{when a participant's behavior implicitly relied on their pre-existing knowledge} of the world and the FLOSS context. Participants exhibited these reliance behaviors when they collaborated by relying on their subconscious understanding of social norms without consciously thinking about the knowledge or putting it into words. For example, we recall the episode in \textit{W8} where P8EUa shared a possible feature: \textit{``I also wonder if this sort of chatbot um ... could be, somehow, hooked into any given app. And so it could be some sort of [open source framework] that a developer adds to their application and sets up on the back end.''} At the conclusion, P8DES asked for clarification: \textit{``So instead of a separate platform, something you can implement in existing platforms?''} P8EUa responded by saying\textit{``Yeah. Although I say this as somebody who doesn't know how difficult, you know ... which one of those is more difficult to do.''} In response, P8DES said that design activities are fun because it's supposed to ``\textit{free}'' people from worrying about stuff like implementation. This episode demonstrates P8EUa's tacit knowledge of the impact resource constraints have in a FLOSS context, thus causing P8EUa to express self-doubt but not articulate the underlying reason further.

On the other hand, practical consciousness was also demonstrated \textbf{when a participant made assumptions that were informed by their practical knowledge} and perceptions of the FLOSS status quo. Participants would use these assumptions as motivation to exercise their power, which typically confirmed the status quo. An example occurred in \textit{W1} when P1EU responded to P1DES's idea for implementing synchronous video calls by saying: \textit{``I don't know if an open-source project could implement a [synchronous] weekly meeting like that. It would be great if they could.''} This demonstrated that P1EU's concerns about this feature were based on assumptions of FLOSS and stakeholder communication behaviors. Additionally, participants made assumptions about each other as they worked together. For example, in \textit{W6}, P6EU said to P6DEV \textit{``Wow, you're really good at this,''} referring to P6DEV's skills using the collaboration tool, Excalidraw, signaling P6EU's (who had never used Excalidraw before the workshop) perspective of P6DEV as an authority of the tool. However, P6DEV responded to the compliment by saying it was actually their first time using the software. In this episode, P6EU assumed P6DEV's skill level based only on a few actions performed by P6DEV, reinforcing a power hierarchy.

\subsection{Discursive Consciousness}
\label{sec:d3_discursive_conscious}
Consciousness becomes discursive when a person directly puts into words what they are thinking; this is achieved by shifting abstract and subconscious thoughts to the conscious mind where thoughts become concrete. Discursive consciousness is viewed as the underlying force for de-structuration (see Section~\ref{sec:power-theory}). Through critical reflection that transforms tacit knowledge into discursive knowledge, people may choose to exercise their power (D-1) to de-structure the social norms (D-2), and, if done consistently over time, a new status quo will form. Because the focus group and design activity were both designed to inspire critical thinking on challenges in FLOSS usability, much of the participants' behavior was already discursive as they actively discussed, brainstormed, and collaborated. For this reason, we focused our analysis on behavior that demonstrated their discursive consciousness through active and vocal \textit{critical thinking and engagement} regarding the status quo of FLOSS usability. We identified three discursive behaviors performed by participants, often resulting in de-structuration.

First, participants actively convinced a teammate through \textbf{articulated logical reasoning about a topic}, either arguing against or for an idea or strategy. \textit{Resource utilization} was occasionally used as an aid, such as design artifacts, personas, external documentation, or past experiences. For instance, in \textit{W8}, P8EUb sought confirmation about their tentative design pitch in the earlier ideation phases (a personalized and easy-to-use chatbot). P8DES responded by justifying the idea, saying that \textit{``there's a reason a lot of companies start using [chatbots]. These days, it's easy. You can just type [your issues].''} The chatbot feature served as a \textit{de-structuration} of the current norms of impersonal communication methods in FLOSS (i.e., issue tracking). In another example, P5DEV and P5DES noticed that there was no persona for a product manager role who coordinates and oversees the development. P5DES reasoned that \textit{``[product managers are] historically not well resourced in open source and I think there's an assumption that these facilitator roles are to be done by the tools.''} P5DEV resonated with this logic about the status quo and considered this role important for ``\textit{coordinating tasks and checking contexts}.'' These reasoning steps evolved into their final pitch, which included features to fulfill such a role.

Second, participants \textbf{voiced empathy towards other stakeholder groups}. These expressions were often made through phrases such as ``\textit{if I'm a [STAKEHOLDER TYPE]...,}'' followed by a statement where the participants put themselves in the shoes of another stakeholder. Participants utilized empathy most frequently to imagine the needs of an end-user or a non-technical individual. For example, in \textit{W1}, P1DEV indicated their agreement to P1DES’s design of a front-end abstraction for a GitHub repository by empathizing with non-technical end-users and added that P1DES's design \textit{``get[s] away from GitHub scaring people off, [makes] it more user friendly.''} Eventually, P1DES's designs were included in the pitch, de-structuring the current development-centric culture of FLOSS. Similarly, in \textit{W7}, the participants were brainstorming on usability-specific features for end-users. P7DES, empathizing with users feeling stressed, suggested a feature to support first-time users: \textit{``Users don't really have to watch a video or read the documentation for that particular tool. There's a walk-through for the first two minutes, and after a sign-up -- well, not sign-up ... because sign up also is kind of stressful for users in terms of experiences.''} The other participants agreed and this was summarized as a key feature of their pitch.

Third, participants \textbf{expressed self-awareness of their reliance on tacit knowledge}, especially as it related to their own biases. In these cases, participants verbally acknowledged their biases and added justifications, often through a lens of the assumptions about other stakeholder groups and the FLOSS status quo. For example, in \textit{W5}, when brainstorming issues experienced by FLOSS stakeholders, P5DEV suggested to P5DES: \textit{``maybe you [should summarize the developer's issues] because that way you won't have that bias.''} This implied that -- as a developer -- if P5DEV summarized the developer persona issues, they might introduce bias originating from their own experiences or opinions. By suggesting that P5DES summarize the developer persona instead, P5DEV demonstrated an awareness of their own tacit information.
Similarly, in an earlier example that occurred in \textit{W2}, when P2DES realized and acknowledged their complacency in assuming that FLOSS developers do not get involved in usability discussions, P2DES said \textit{``I'm kind of falling into my own thing ... that I complain about designers, where they don't involve the developers until, like, the very end.''} P2DES later changed their mindset, ultimately saying that \textit{``if [developers] want to advocate for usability changes, they need to be involved earlier on.''} By reframing their tacit understanding of how developers ``\textit{usually}'' act to the ``\textit{ideal}'' way developers could act, the participants de-structured the status quo.
\section{Discussion}
\label{sec:discussion}
\begin{figure*}[t]
    \centering
    \includegraphics[width=\textwidth]{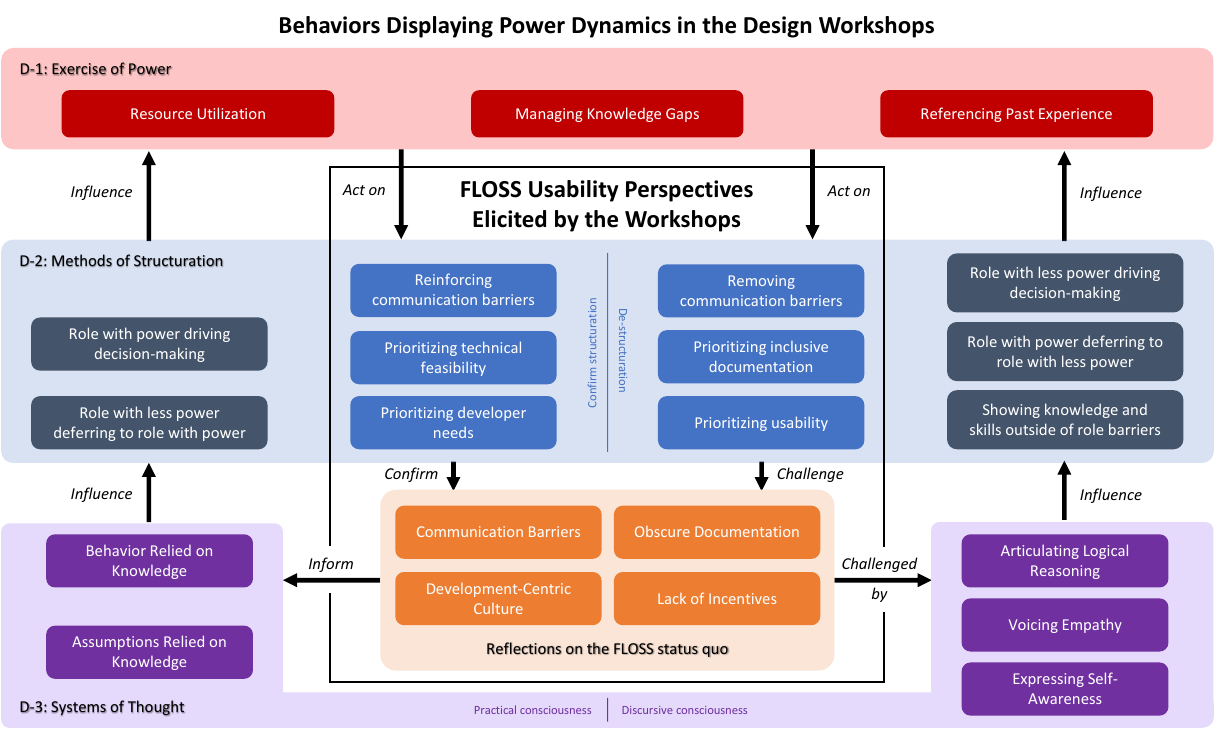}
    \caption{Synthesis of our study results, demonstrating how the design workshops unveiled and mediated the power dynamics of FLOSS.}
    \Description{Four components - (1) D-1 Exercise of Power, (2) D-2 Methods of Structuration, (3) D-3 Systems of Thought, (4) Reflections on the FLOSS Status Quo. Reflections on the FLOSS Status Quo and the part of D-2 Methods of Structuration related to design ideas belong to the aspect of FLOSS Usability Perspectives Elicited by Workshops. The rest belongs to the aspect of Behaviors Displaying Power Dynamics in the Design Workshops. D-1 Exercise of Power contains resource utilization, managing knowledge gaps, and referencing past experience. D-1 acts on the D-2 Methods of Structuration which contain two processes: confirm structuration and de-structuration. Confirm structuration contains reinforcing communication barriers, prioritizing technical feasibility, prioritizing developer needs, role with power driving decision-making, and role with less power deferring to the role with power. De-structuration involves removing communication barriers, prioritizing inclusive documentation, prioritizing usability, role with less power driving decision-making, role with power deferring to role with less power, and showing knowledge skills outside of role barriers. Confirm-structuration confirms the status quo and de-structuration challenges the status quo. There are four reflections contained in the status quo: communication barriers, obscure documentation, development-centric culture, and lack of incentives. D-3 Systems of Thought contain practical consciousness and discursive consciousness. The status quo informs the practical consciousness, which contains behavior relied on knowledge and assumptions relied on knowledge. Discursive consciousness challenges the status quo and contains articulating logical reasoning, voicing empathy, and expressing self-awareness. D-3 Systems of Thought influence the D-2 Methods of Structuration which influence D-1 Exercises of Power.}
    \label{fig:results_summary}
\end{figure*}

Through our focus groups and design workshops, we exposed and observed factors contributing to role-based power dynamics during direct collaboration among different FLOSS stakeholders.
The focus group and workshop artifacts (i.e., personas) stimulated discursive thinking and critical conversations about the status quo of FLOSS usability. The design workshops unveiled and amplified the collaborative processes and encouraged further discursive conversations that lead to exercises of power, either reinforcing or challenging the status quo. Our findings should not be viewed in isolation, but rather as cycles wherein reflection (D-3) and action (D-1) on undesired role-based power structures can lead to changes in social structures (D-2). 
Figure~\ref{fig:results_summary} depicts the two cycles that surfaced through communication and reflection during the design workshops, either confirming or challenging the existing FLOSS structure. As different aspects of power are intricately entangled, improving FLOSS usability needs a holistic approach that taps into each dimension of power simultaneously.

Well-established strategies exist, which might empower the end-users, such as adopting usability best practices and widely distributing relevant resources. However, our study, along with prior work, has shown that those strategies often fall short and require developer buy-in~\cite{hellman_facilitating_2021, geiger_2021}. Continuous implementation of those strategies stems from realizing the limitation of the current developer-driven FLOSS structure and initializing a concentrated push to de-structurize the status quo. Reflecting on possible ways to leverage our power dimension's results, we discuss the implications of our study along each power dimension and propose concrete design recommendations and suggestions for future work to potentially raise the discursive consciousness and initiate the de-structuration cycle.

\subsection{Equalizing Exercise of Power Among Stakeholders.}
Active participation in FLOSS is built on effective exercises of power and participation, one of the core tenets of FLOSS. However, the value of ``openness'' in the FLOSS paradigm assumes that diverse roles can participate equally in those projects, allowing for adequate capacities to exercise power. Our analysis reveals three major ways of exercising power adopted by our participants: \textit{resource utilization}, \textit{managing knowledge gaps}, and \textit{referencing past experience}. Therefore, we should consider ways such exercises of power can be leveraged to work towards discursive awareness and trigger de-structuration. It is worth noting that existing FLOSS development platforms, such as GitHub, already support all three aspects of exercising power by the developers (e.g., public repositories, standardized development documentation, code search and code review, etc.), but similar support for the other roles (e.g., designers and end-users) is limited, non-existent, or detached from the FLOSS workflow~\cite{hellman_facilitating_2021, Hellman_Chen_Uddin_Cheng_Guo_2022, rajanen_power_2015}. We argue that facilitating all FLOSS stakeholders to exercise their power through these three ways can effectively enable participation. 

In particular, we propose that FLOSS resources, such as documents, tools, and design and development artifacts, should be easily retrievable and understood by all stakeholders. This is especially useful for smaller FLOSS projects, in which resources are less voluminous and specialized. In fact, many of the design pitches (Section~\ref{sec:design-pitches}) proposed by our participant suggested means to address this aspect in the D-1 power dimension. For example, four workshops (W1, W4, W7, and W8) proposed designs to support more inclusive documentation through simplifying landing pages, highlighting the location of documentation or resource links, having dedicated education portals for diverse stakeholders, and even end-user-targeted chatbots that can help provide information about the FLOSS project. 
Automated solutions that can be customized to meet different stakeholders' needs, such as chatbots, might also help bridge the \textit{knowledge gap} among FLOSS stakeholders. Indeed, individuals might be more willing to ask questions to non-human agents that they wouldn't ask otherwise for fear of judgment, embarrassment, fears of offending developers, or wasting people's time~\cite{geiger_2021}. At the same time, however, chatbots might lead to further separation between stakeholders by removing human contact, therefore further aggravating confirm-structuration behaviors, especially \textit{inhibition behaviors} (see Section~\ref{sec:d1-knowledge-gap}). Further work should explore the impact of automated tools on FLOSS communities when navigating this social-technical space (e.g., similar to the work by \citet{SmithYSHTZ20}), as well as the interplay among AI and automation, de-structure knowledge, and communication norms. 

Maintaining educational resources and documentation for all stakeholders might become less feasible as projects scale and expertise beyond development needs to be integrated~\cite{geiger_2021}, especially for complex FLOSS projects. Some of our workshop participants addressed this tension by designing features highlighting the need for volunteers with niche skills, such as technical writers and project managers, to diversify the community. This effort can be further supplemented by previous work on why people contribute voluntarily~\cite{AryaGR24} and explore ways to meet their goals~\cite{CobbMPBCDS14}. In conjunction, our results indicated that we should support initiatives for all stakeholders of a FLOSS project to bond both practically and mentally to facilitate a collaborative growth mindset. FLOSS teams should maintain inclusive language (e.g., gender inclusivity, expertise inclusivity, etc.), use familiar tools (e.g., Discord, etc.), require low commitment, and may consider forming an ``onboarding committee'' to greet, encourage, and validate newcomers, regardless of their technical knowledge. In doing so, FLOSS projects can foster an inclusive, curious community while also managing diverse stakeholders' expectations, such as those of the community maintainers~\cite{geiger_2021}, both new and existing developers~\cite{santos2022}, as well as designers and end users~\cite{hellman_facilitating_2021}.

Furthermore, integration of tools that facilitate capturing and sharing of \textit{personal and in-situ experiences with FLOSS} (e.g., experiences of designing, developing, and using FLOSS) would provide valuable context when discussing usability, especially benefiting FLOSS projects with a large and diverse user base. This would in turn support negotiation and argumentation about usability issues~\cite{wang_argulens_2020, AjmaniFTPGD24}. 
For example, dedicated experience-sharing forums (e.g., for end-users of a FLOSS project~\cite{Hellman_Chen_Uddin_Cheng_Guo_2022}) should consider ways to support more effective discussion and brainstorming based on the in-situ experience of software usage. By including features specifically aimed to inspire empathy, check assumptions, and prompt self-awareness in such forums (e.g., incorporating suggestions made in previous work by \citet{wang_argulens_2020,filippova_2015,jamieson_2022}), a safe space can be created for value-based conflict to challenge the norms of a project. 
At the same time, such forums should maintain a close integration with developer-centric platforms such as GitHub to avoid segregation of FLOSS stakeholder roles~\cite{hellman_facilitating_2021}. 

\subsection{Striving for Changes in the Social Structure of FLOSS.}
Existing work on improving FLOSS usability has proposed to engage core users for feedback~\cite{terry_perceptions_2010} or to integrate the usability experts into the development team~\cite{rajanen_power_2015}. In both cases, however, the decision power is still mainly controlled by the developers, an ongoing reality we see reflected in our confirm-structuration results. While previous work by \citet{wubishet2013participation} has suggested addressing unbalanced decision power through adopting collaborative design activities for FLOSS through user-developer workshops and brainstorming sessions, there was limited impact of those activities on usability. They observed a ``lack of real end-user participation'' and stated that the ``re-balancing of power relationships by the movement does not seem to have fully materialised''~\cite{wubishet2013participation}. Indeed, adopting participation-based activities without properly revising the decision-making structures or considering support for the less-powered roles is not likely to introduce substantial and long-term change~\cite{rajanen_power_2015,bach2010, nichols_usability_2003}. Therefore, FLOSS communities should restructure their decision-making process to enable the end-users' and designers' access to decision power for a project, ranging from individual features of the software to the future directions of the project and values of community~\cite{wang_how_2020, Iivari09}.
These efforts, combined with implementing changes to equalize the potential to exercise power, can allow FLOSS projects to work towards meaningful changes.

However, such fundamental restructuring remains challenging, especially for mature FLOSS projects, in which established contribution patterns and ingrained workflows may create significant barriers to change. This was also reflected in some design pitches. For example, W2's design pitch proposed a crucial feature to make a designer, instead of a developer, decide the priority of usability issues and automatically notify developers if an issue needs to be addressed. Here, the designer decides what is a worthwhile usability issue, yet the developer still has the final decision power and end-users remain with little decision power. W6 and W7 both proposed streamlined communication mechanisms among FLOSS stakeholders through synchronous conferencing. It is interesting to note the potential of synchronous communications in disrupting current decision structures~\cite{geiger_2021}; however, it might risk further burn-out and over-inundating developers~\cite{filippova_2015}. To find effective and sustainable decision-making alternatives, we call for future work to explore different modes of governance for FLOSS and to understand their impact on democratization. Previous work from CSCW and HCI provides valuable insights. For example, efforts can be devoted to providing resources and fostering a culture that is primarily driven by end-users (i.e., the end-users act directly as designers and makers of the technology) so that the communities and the software they make would manifest the end-users' values and advocate for inclusion~\cite{VyasKC23, FoxUR15}.

At the same time, empathy-building tools such as personas can play an important role in social structure change. Reflecting on our personas, the decision power structures were not explicitly depicted; doing so would likely have inspired more discussions on decision power structures during the workshops. Building on our result in D2, we expect to update and create more full-fledged personas to reflect stakeholders' challenges on decision power for future use. These tools, similar to the work of GenderMag~\cite{Burnett2016GenderMag} that used personas for gender inclusiveness considerations in software development, would further improve the decision-making structure of a wide range of FLOSS projects by stressing the needs of the less-represented roles in the current FLOSS tools. This will help encourage value-charged discussions~\cite{jamieson_2022} through the sharing of knowledge from both research and practices to form a larger de-structuration cycle that challenges the status quo and ultimately leads to new social norms.

\subsection{Raising Discursive Consciousness Through Collective Reflection.}
In our study, the focus groups ensured a large-scale demonstration of discursive behaviors even before any participants started the design activities. This was critical to allow participants to concretely think about and discuss any issues that had been in their practical consciousness prior to the workshop; it encouraged both \textit{empathy} and \textit{self-awareness} of their assumptions towards other roles. Once the knowledge became discursive, participants could readily identify the issues they wanted to address in the design activity, thus influencing the pitch outcomes. For example, \textit{W1} participants discursively reflected on \textit{obscure documentation} during the focus group and its negative effects on participation. Through this reflection, P1EU and P1DES expressed empathy for non-technical end-users becoming overwhelmed during initial GitHub encounters. As a result, the \textit{W1} design pitch included a user-friendly centralized platform, separate from GitHub, for hosting FLOSS projects in their final design pitch. 

This and other examples from our study demonstrated the benefits of conducting intentional conversations, which include increasing empathy, promoting a deeper understanding of usability problems, and encouraging an appreciation for other stakeholders' experiences; those findings align with important factors identified in previous work to improve FLOSS usability~\cite{hellman_facilitating_2021,jamieson_2022, SaneiC24}. Such conversations can be held within any individual FLOSS project or, more beneficially, in spaces where cross-project discussions can take place. Notably, these types of conversations also echo and add nuances to the findings by~\citet{filippova_2015} regarding the importance of handling normative conflicts (i.e., discussions on the discrepancy between the stated aims of a group and the actual practices) to achieve long-term FLOSS project sustainability. Therefore, we believe dedicated platforms or opportunities that enable all stakeholders to collectively reflect on the status quo of the FLOSS development process can greatly influence its usability. Similar to the focus group stage of our methodology, such reflection can be the first step for FLOSS communities to develop discursive knowledge and foster self-examination and empathy. Efforts to extend existing initiatives like Open Source Design\footnote{https://opensourcedesign.net}) to the broader FLOSS community should be advocated. 

On the other hand, our results also underscored the role of \textit{assumption making} in the confirming cycles. Assumptions and behaviors based on existing knowledge can be hard to break, especially when they are well-accepted or even expected notions, such as ``developers are the ones in charge of deciding and implementing the changes to the FLOSS.'' Speculative approaches~\cite{ChopraCCHDV22, ProstMT15, Dunne2013} that encourage questioning practical knowledge are valuable tools to help identify those assumptions and, therefore, shift the practical consciousness to discursive. By intentionally creating an environment that requires stakeholders to push the boundaries of reality, we can interrogate the very definition of FLOSS, the social and technical assumptions that support the definition, as well as the relationship between usability, participation, and power in a more fundamental manner. The outcome of those speculative approaches might lead to interventions that disrupt a person's reliance on their practical knowledge, such as tools that allow stakeholders to easily check their communications for potential bias and mediate the typically invisible influence of assumptions. 

\subsection{Limitations}
Several main limitations arise from our recruitment, the persona documents, and the design task. First of all, our study focused on analyzing power dynamics among different FLOSS roles. While this scope provides us valuable information for empowering designers and end-users in making impacts to improve FLOSS usability, we acknowledge that power imbalances caused by other factors such as gender, race, and language exist in FLOSS communities. Although our participants were diverse in terms of race, nationality, and culture, the gender ratio was skewed towards males. Related to the personas provided to the participants, first, their gender identities were not explicitly stated, which might cause participants to assume gender. Additionally, the end-user and designer personas were people of color, while the developer persona was presented as white. This possibly influenced participants' sub-conscious perceptions about the stakeholders, including the power dynamics among the personas themselves. Future studies could extend this investigation to explore the relationship between stakeholder demographics (e.g., gender and race) and power dynamics in FLOSS usability. 

Further, our design workshops were conceived to provide a feasible proxy of how roles in FLOSS community negotiate with others to achieve a common objective (see Section~\ref{sec:methods_reflection}). However, these workshops can only partially represent the complexity of real-world FLOSS communities when addressing usability, in which a larger community (rather than a few participants as in our workshops) is involved. Moreover, the design workshops as a research method could only capture short-term collaboration issues. Thus, future longitudinal studies that involve a wide spectrum of FLOSS community members are needed, although we acknowledge the challenges of designing such studies to make the hidden power dynamics visible.

\section{Conclusion}
We present a study of collaborative design in a FLOSS context to understand the interplay of power dynamics and participation among developers, designers, and end-users. In particular, we conducted eight design workshops with key FLOSS usability stakeholders (i.e., developers, designers, and end users) to understand how they exercise power during collaboration (D-1), the effects of their behaviors on the status quo (D-2), and how their underlying thought processes were demonstrated (D-3). Our study revealed that the typical role-based power dynamics in FLOSS (i.e., developers have power over designers and end-users) can be replicated during face-to-face collaboration due to a reliance on tacit knowledge of the status quo. However, we also found that participants exercised their power to challenge the status quo when they instilled more self-awareness and developed empathy toward other stakeholder roles, and when those traditionally with less power were engaged during collaboration. Our work contributes to a close examination of power dynamics among the main FLOSS stakeholder roles, suggesting ways to build a more democratic process to improve FLOSS usability. Our study also represents an exemplar of using collaborative design workshops as a research method for investigating and mediating power dynamics that are usually hidden in real-world settings.

\begin{acks}
    We thank our participants for their time and valuable insights. We also thank the anonymous reviewers for helping us improve the paper. This work is partially supported by the Alfred P. Sloan Foundation (G-2021-16745).
\end{acks}

\bibliographystyle{ACM-Reference-Format}
\bibliography{ref}


\begin{thebibliography}{67}


\ifx \showCODEN    \undefined \def \showCODEN     #1{\unskip}     \fi
\ifx \showISBNx    \undefined \def \showISBNx     #1{\unskip}     \fi
\ifx \showISBNxiii \undefined \def \showISBNxiii  #1{\unskip}     \fi
\ifx \showISSN     \undefined \def \showISSN      #1{\unskip}     \fi
\ifx \showLCCN     \undefined \def \showLCCN      #1{\unskip}     \fi
\ifx \shownote     \undefined \def \shownote      #1{#1}          \fi
\ifx \showarticletitle \undefined \def \showarticletitle #1{#1}   \fi
\ifx \showURL      \undefined \def \showURL       {\relax}        \fi
\providecommand\bibfield[2]{#2}
\providecommand\bibinfo[2]{#2}
\providecommand\natexlab[1]{#1}
\providecommand\showeprint[2][]{arXiv:#2}

\bibitem[Ajmani et~al\mbox{.}(2024)]%
        {AjmaniFTPGD24}
\bibfield{author}{\bibinfo{person}{Leah~Hope Ajmani}, \bibinfo{person}{Jasmine~C. Foriest}, \bibinfo{person}{Jordan Taylor}, \bibinfo{person}{Kyle Pittman}, \bibinfo{person}{Sarah~A. Gilbert}, {and} \bibinfo{person}{Michael~Ann DeVito}.} \bibinfo{year}{2024}\natexlab{}.
\newblock \showarticletitle{Whose {Knowledge} is {Valued}? {Epistemic} {Injustice} in {CSCW} {Applications}}.
\newblock \bibinfo{journal}{\emph{Proc. ACM Hum. Comput. Interact.}} \bibinfo{volume}{8}, \bibinfo{number}{CSCW2} (\bibinfo{year}{2024}), \bibinfo{pages}{1--28}.
\newblock
\href{https://doi.org/10.1145/3687062}{doi:\nolinkurl{10.1145/3687062}}


\bibitem[Allen(2018)]%
        {Allen_2018}
\bibfield{author}{\bibinfo{person}{Amy Allen}.} \bibinfo{year}{2018}\natexlab{}.
\newblock \bibinfo{booktitle}{\emph{The {Power} of {Feminist} {Theory}}}.
\newblock \bibinfo{publisher}{Routledge}, \bibinfo{address}{New York}.
\newblock
\showISBNx{978-0-429-49593-9}
\href{https://doi.org/10.4324/9780429495939}{doi:\nolinkurl{10.4324/9780429495939}}


\bibitem[Andreasen et~al\mbox{.}(2006)]%
        {andreasen_usability_2006}
\bibfield{author}{\bibinfo{person}{Morten~Sieker Andreasen}, \bibinfo{person}{Henrik~Villemann Andreasen}, \bibinfo{person}{Simon~Ormholt Schrøder}, {and} \bibinfo{person}{Jan Stage}.} \bibinfo{year}{2006}\natexlab{}.
\newblock \showarticletitle{Usability in {Open} {Source} {Software} {Development}: {Opinions} and {Practice}}.
\newblock \bibinfo{journal}{\emph{Information Technology and Control}} \bibinfo{volume}{35}, \bibinfo{number}{3} (\bibinfo{year}{2006}), \bibinfo{pages}{303--312}.
\newblock
\href{https://doi.org/10.5755/j01.itc.35.3.11776}{doi:\nolinkurl{10.5755/j01.itc.35.3.11776}}


\bibitem[Arya et~al\mbox{.}(2024)]%
        {AryaGR24}
\bibfield{author}{\bibinfo{person}{Deeksha~M. Arya}, \bibinfo{person}{Jin L.~C. Guo}, {and} \bibinfo{person}{Martin~P. Robillard}.} \bibinfo{year}{2024}\natexlab{}.
\newblock \showarticletitle{Why People Contribute Software Documentation}. In \bibinfo{booktitle}{\emph{Proceedings of the 2024 IEEE/ACM 17th International Conference on Cooperative and Human Aspects of Software Engineering}} \emph{(\bibinfo{series}{CHASE '24})}. \bibinfo{publisher}{ACM}, \bibinfo{address}{New York, USA}, \bibinfo{pages}{91–96}.
\newblock
\showISBNx{9798400705335}
\href{https://doi.org/10.1145/3641822.3641881}{doi:\nolinkurl{10.1145/3641822.3641881}}


\bibitem[Bach et~al\mbox{.}(2009)]%
        {bach_designers_2009}
\bibfield{author}{\bibinfo{person}{Paula~M. Bach}, \bibinfo{person}{Robert DeLine}, {and} \bibinfo{person}{John~M. Carroll}.} \bibinfo{year}{2009}\natexlab{}.
\newblock \showarticletitle{Designers wanted: participation and the user experience in open source software development}. In \bibinfo{booktitle}{\emph{Proceedings of the 27th International Conference on Human Factors in Computing Systems, ({CHI} 2009)}}. \bibinfo{publisher}{{ACM}}, \bibinfo{address}{New York, USA}, \bibinfo{pages}{985--994}.
\newblock
\href{https://doi.org/10.1145/1518701.1518852}{doi:\nolinkurl{10.1145/1518701.1518852}}


\bibitem[Bach and Twidale(2010)]%
        {bach2010}
\bibfield{author}{\bibinfo{person}{Paula~M. Bach} {and} \bibinfo{person}{Michael~B. Twidale}.} \bibinfo{year}{2010}\natexlab{}.
\newblock \showarticletitle{Social participation in open source: what it means for designers}.
\newblock \bibinfo{journal}{\emph{Interactions}} \bibinfo{volume}{17}, \bibinfo{number}{3} (\bibinfo{year}{2010}), \bibinfo{pages}{70--74}.
\newblock
\href{https://doi.org/10.1145/1744161.1744177}{doi:\nolinkurl{10.1145/1744161.1744177}}


\bibitem[Barcellini et~al\mbox{.}(2014)]%
        {barcellini_situated_2014}
\bibfield{author}{\bibinfo{person}{Flore Barcellini}, \bibinfo{person}{Françoise Détienne}, {and} \bibinfo{person}{Jean-Marie Burkhardt}.} \bibinfo{year}{2014}\natexlab{}.
\newblock \showarticletitle{A {Situated} {Approach} of {Roles} and {Participation} in {Open} {Source} {Software} {Communities}}.
\newblock \bibinfo{journal}{\emph{Hum. Comput. Interact.}} \bibinfo{volume}{29}, \bibinfo{number}{3} (\bibinfo{year}{2014}), \bibinfo{pages}{205--255}.
\newblock
\href{https://doi.org/10.1080/07370024.2013.812409}{doi:\nolinkurl{10.1080/07370024.2013.812409}}


\bibitem[Bratteteig and Wagner(2016)]%
        {Bratteteig_Wagner_2016}
\bibfield{author}{\bibinfo{person}{Tone Bratteteig} {and} \bibinfo{person}{Ina Wagner}.} \bibinfo{year}{2016}\natexlab{}.
\newblock \showarticletitle{Unpacking the Notion of Participation in Participatory Design}.
\newblock \bibinfo{journal}{\emph{Computer Supported Cooperative Work (CSCW)}} \bibinfo{volume}{25}, \bibinfo{number}{6} (\bibinfo{year}{2016}), \bibinfo{pages}{425–475}.
\newblock
\showISSN{1573-7551}
\href{https://doi.org/10.1007/s10606-016-9259-4}{doi:\nolinkurl{10.1007/s10606-016-9259-4}}


\bibitem[Burnett et~al\mbox{.}(2016)]%
        {Burnett2016GenderMag}
\bibfield{author}{\bibinfo{person}{Margaret Burnett}, \bibinfo{person}{Anicia Peters}, \bibinfo{person}{Charles Hill}, {and} \bibinfo{person}{Noha Elarief}.} \bibinfo{year}{2016}\natexlab{}.
\newblock \showarticletitle{Finding Gender-Inclusiveness Software Issues with GenderMag: A Field Investigation}. In \bibinfo{booktitle}{\emph{Proceedings of the 2016 CHI Conference on Human Factors in Computing Systems}} \emph{(\bibinfo{series}{CHI '16})}. \bibinfo{publisher}{ACM}, \bibinfo{address}{New York, USA}, \bibinfo{pages}{2586–2598}.
\newblock
\showISBNx{9781450333627}
\href{https://doi.org/10.1145/2858036.2858274}{doi:\nolinkurl{10.1145/2858036.2858274}}


\bibitem[Cheng and Guo(2018)]%
        {cheng_how_2018}
\bibfield{author}{\bibinfo{person}{Jinghui Cheng} {and} \bibinfo{person}{Jin L.~C. Guo}.} \bibinfo{year}{2018}\natexlab{}.
\newblock \showarticletitle{How {Do} the {Open} {Source} {Communities} {Address} {Usability} and {UX} {Issues}?: {An} {Exploratory} {Study}}. In \bibinfo{booktitle}{\emph{Extended {Abstracts} of the 2018 {CHI} {Conference} on {Human} {Factors} in {Computing} {Systems}}} \emph{(\bibinfo{series}{{CHI} {EA} '18})}. \bibinfo{publisher}{ACM}, \bibinfo{address}{New York, USA}, \bibinfo{pages}{1--6}.
\newblock
\href{https://doi.org/10.1145/3170427.3188467}{doi:\nolinkurl{10.1145/3170427.3188467}}


\bibitem[Cheng and Guo(2019)]%
        {Cheng2019-roles}
\bibfield{author}{\bibinfo{person}{Jinghui Cheng} {and} \bibinfo{person}{Jin L.~C. Guo}.} \bibinfo{year}{2019}\natexlab{}.
\newblock \showarticletitle{Activity-based analysis of open source software contributors: roles and dynamics}. In \bibinfo{booktitle}{\emph{Proceedings of the 12th International Workshop on Cooperative and Human Aspects of Software Engineering, CHASE@ICSE 2019}}. \bibinfo{publisher}{{IEEE} / {ACM}}, \bibinfo{address}{New York, USA}, \bibinfo{pages}{11--18}.
\newblock
\href{https://doi.org/10.1109/CHASE.2019.00011}{doi:\nolinkurl{10.1109/CHASE.2019.00011}}


\bibitem[Chopra et~al\mbox{.}(2022)]%
        {ChopraCCHDV22}
\bibfield{author}{\bibinfo{person}{Simran Chopra}, \bibinfo{person}{Rachel~E Clarke}, \bibinfo{person}{Adrian~K Clear}, \bibinfo{person}{Sara Heitlinger}, \bibinfo{person}{Ozge Dilaver}, {and} \bibinfo{person}{Christina Vasiliou}.} \bibinfo{year}{2022}\natexlab{}.
\newblock \showarticletitle{Negotiating sustainable futures in communities through participatory speculative design and experiments in living}. In \bibinfo{booktitle}{\emph{Proceedings of the 2022 CHI Conference on Human Factors in Computing Systems}} \emph{(\bibinfo{series}{CHI '22})}. \bibinfo{publisher}{ACM}, \bibinfo{address}{New York, USA}, Article \bibinfo{articleno}{334}, \bibinfo{numpages}{17}~pages.
\newblock
\showISBNx{9781450391573}
\href{https://doi.org/10.1145/3491102.3501929}{doi:\nolinkurl{10.1145/3491102.3501929}}


\bibitem[Cobb et~al\mbox{.}(2014)]%
        {CobbMPBCDS14}
\bibfield{author}{\bibinfo{person}{Camille Cobb}, \bibinfo{person}{Ted McCarthy}, \bibinfo{person}{Annuska Perkins}, \bibinfo{person}{Ankitha Bharadwaj}, \bibinfo{person}{Jared Comis}, \bibinfo{person}{Brian Do}, {and} \bibinfo{person}{Kate Starbird}.} \bibinfo{year}{2014}\natexlab{}.
\newblock \showarticletitle{Designing for the deluge: understanding \& supporting the distributed, collaborative work of crisis volunteers}. In \bibinfo{booktitle}{\emph{Proceedings of the 17th ACM Conference on Computer Supported Cooperative Work \& Social Computing}} \emph{(\bibinfo{series}{CSCW '14})}. \bibinfo{publisher}{ACM}, \bibinfo{address}{New York, USA}, \bibinfo{pages}{888–899}.
\newblock
\showISBNx{9781450325400}
\href{https://doi.org/10.1145/2531602.2531712}{doi:\nolinkurl{10.1145/2531602.2531712}}


\bibitem[Crowston and Howison(2005)]%
        {crowston2005social}
\bibfield{author}{\bibinfo{person}{Kevin Crowston} {and} \bibinfo{person}{James Howison}.} \bibinfo{year}{2005}\natexlab{}.
\newblock \showarticletitle{The social structure of free and open source software development}.
\newblock \bibinfo{journal}{\emph{First Monday}} \bibinfo{volume}{10}, \bibinfo{number}{2} (\bibinfo{year}{2005}).
\newblock
\href{https://doi.org/10.5210/FM.V10I2.1207}{doi:\nolinkurl{10.5210/FM.V10I2.1207}}


\bibitem[Dabbish et~al\mbox{.}(2012)]%
        {dabbish_social_2012}
\bibfield{author}{\bibinfo{person}{Laura Dabbish}, \bibinfo{person}{Colleen Stuart}, \bibinfo{person}{Jason Tsay}, {and} \bibinfo{person}{Jim Herbsleb}.} \bibinfo{year}{2012}\natexlab{}.
\newblock \showarticletitle{Social coding in GitHub: transparency and collaboration in an open software repository}. In \bibinfo{booktitle}{\emph{Proceedings of the ACM 2012 Conference on Computer Supported Cooperative Work}} \emph{(\bibinfo{series}{CSCW '12})}. \bibinfo{publisher}{ACM}, \bibinfo{address}{New York, USA}, \bibinfo{pages}{1277–1286}.
\newblock
\showISBNx{9781450310864}
\href{https://doi.org/10.1145/2145204.2145396}{doi:\nolinkurl{10.1145/2145204.2145396}}


\bibitem[Dahl(1957)]%
        {Dahl_1957}
\bibfield{author}{\bibinfo{person}{Robert~A. Dahl}.} \bibinfo{year}{1957}\natexlab{}.
\newblock \showarticletitle{The concept of power}.
\newblock \bibinfo{journal}{\emph{Behavioral Science}} \bibinfo{volume}{2}, \bibinfo{number}{3} (\bibinfo{year}{1957}), \bibinfo{pages}{201--215}.
\newblock
\showISSN{1099-1743}
\href{https://doi.org/10.1002/bs.3830020303}{doi:\nolinkurl{10.1002/bs.3830020303}}


\bibitem[Dawood et~al\mbox{.}(2019)]%
        {dawood_mapping_2019}
\bibfield{author}{\bibinfo{person}{Kareem~Abbas Dawood}, \bibinfo{person}{Khaironi~Yatim Sharif}, \bibinfo{person}{A.~A. Zaidan}, \bibinfo{person}{Abdul Azim~Abdul Ghani}, \bibinfo{person}{Hazura Zulzalil}, {and} \bibinfo{person}{B.~B. Zaidan}.} \bibinfo{year}{2019}\natexlab{}.
\newblock \showarticletitle{Mapping and Analysis of Open Source Software {(OSS)} Usability for Sustainable {OSS} Product}.
\newblock \bibinfo{journal}{\emph{{IEEE} Access}}  \bibinfo{volume}{7} (\bibinfo{year}{2019}), \bibinfo{pages}{65913--65933}.
\newblock
\href{https://doi.org/10.1109/ACCESS.2019.2914368}{doi:\nolinkurl{10.1109/ACCESS.2019.2914368}}


\bibitem[Dunne and Raby(2013)]%
        {Dunne2013}
\bibfield{author}{\bibinfo{person}{Anthony Dunne} {and} \bibinfo{person}{Fiona Raby}.} \bibinfo{year}{2013}\natexlab{}.
\newblock \bibinfo{booktitle}{\emph{Speculative {Everything}: {Design}, {Fiction}, and {Social} {Dreaming}}}.
\newblock \bibinfo{publisher}{The MIT Press}, \bibinfo{address}{Cambridge, MA, USA}.
\newblock
\showISBNx{978-0-262-01984-2}


\bibitem[Farr(2018)]%
        {Farr_2018}
\bibfield{author}{\bibinfo{person}{Michelle Farr}.} \bibinfo{year}{2018}\natexlab{}.
\newblock \showarticletitle{Power dynamics and collaborative mechanisms in co-production and co-design processes}.
\newblock \bibinfo{journal}{\emph{Critical Social Policy}} \bibinfo{volume}{38}, \bibinfo{number}{4} (\bibinfo{year}{2018}), \bibinfo{pages}{623--644}.
\newblock
\href{https://doi.org/10.1177/0261018317747444}{doi:\nolinkurl{10.1177/0261018317747444}}


\bibitem[Feller and Fitzgerald(2000)]%
        {feller_framework_2000}
\bibfield{author}{\bibinfo{person}{Joseph Feller} {and} \bibinfo{person}{Brian Fitzgerald}.} \bibinfo{year}{2000}\natexlab{}.
\newblock \showarticletitle{A framework analysis of the open source software development paradigm}. In \bibinfo{booktitle}{\emph{Proceedings of the Twenty-First International Conference on Information Systems, ({ICIS} 2000)}}, \bibfield{editor}{\bibinfo{person}{Soon Ang}, \bibinfo{person}{Helmut Krcmar}, \bibinfo{person}{Wanda~J. Orlikowski}, \bibinfo{person}{Peter Weill}, {and} \bibinfo{person}{Janice~I. DeGross}} (Eds.). \bibinfo{publisher}{Association for Information Systems}, \bibinfo{pages}{58--69}.
\newblock
\urldef\tempurl%
\url{http://aisel.aisnet.org/icis2000/7}
\showURL{%
\tempurl}


\bibitem[Feller et~al\mbox{.}(2005)]%
        {feller_perspectives_2005}
\bibfield{author}{\bibinfo{person}{Joseph Feller}, \bibinfo{person}{Brian Fitzgerald}, \bibinfo{person}{Scott~A Hissam}, {and} \bibinfo{person}{Karim~R Lakhani}.} \bibinfo{year}{2005}\natexlab{}.
\newblock \bibinfo{booktitle}{\emph{Perspectives on free and open source software}}.
\newblock \bibinfo{publisher}{MIT Press}, \bibinfo{address}{Cambridge, Mass.}
\newblock
\href{https://doi.org/10.7551/mitpress/5326.001.0001}{doi:\nolinkurl{10.7551/mitpress/5326.001.0001}}


\bibitem[Filippova and Cho(2015)]%
        {filippova_2015}
\bibfield{author}{\bibinfo{person}{Anna Filippova} {and} \bibinfo{person}{Hichang Cho}.} \bibinfo{year}{2015}\natexlab{}.
\newblock \showarticletitle{Mudslinging and Manners: Unpacking Conflict in Free and Open Source Software}. In \bibinfo{booktitle}{\emph{Proceedings of the 18th {ACM} Conference on Computer Supported Cooperative Work {\&} Social Computing, ({CSCW} 2015)}}. \bibinfo{publisher}{{ACM}}, \bibinfo{address}{New York, USA}, \bibinfo{pages}{1393--1403}.
\newblock
\href{https://doi.org/10.1145/2675133.2675254}{doi:\nolinkurl{10.1145/2675133.2675254}}


\bibitem[Foucault(1982)]%
        {foucault1982subject}
\bibfield{author}{\bibinfo{person}{Michel Foucault}.} \bibinfo{year}{1982}\natexlab{}.
\newblock \showarticletitle{The subject and power}.
\newblock \bibinfo{journal}{\emph{Critical inquiry}} \bibinfo{volume}{8}, \bibinfo{number}{4} (\bibinfo{year}{1982}), \bibinfo{pages}{777--795}.
\newblock


\bibitem[Fox et~al\mbox{.}(2015)]%
        {FoxUR15}
\bibfield{author}{\bibinfo{person}{Sarah Fox}, \bibinfo{person}{Rachel~Rose Ulgado}, {and} \bibinfo{person}{Daniela Rosner}.} \bibinfo{year}{2015}\natexlab{}.
\newblock \showarticletitle{Hacking Culture, Not Devices: Access and Recognition in Feminist Hackerspaces}. In \bibinfo{booktitle}{\emph{Proceedings of the 18th ACM Conference on Computer Supported Cooperative Work \& Social Computing}} \emph{(\bibinfo{series}{CSCW '15})}. \bibinfo{publisher}{ACM}, \bibinfo{address}{New York, USA}, \bibinfo{pages}{56–68}.
\newblock
\showISBNx{9781450329224}
\href{https://doi.org/10.1145/2675133.2675223}{doi:\nolinkurl{10.1145/2675133.2675223}}


\bibitem[Geiger et~al\mbox{.}(2021)]%
        {geiger_2021}
\bibfield{author}{\bibinfo{person}{R.~Stuart Geiger}, \bibinfo{person}{Dorothy Howard}, {and} \bibinfo{person}{Lilly Irani}.} \bibinfo{year}{2021}\natexlab{}.
\newblock \showarticletitle{The Labor of Maintaining and Scaling Free and Open-Source Software Projects}.
\newblock \bibinfo{journal}{\emph{Proc. {ACM} Hum. Comput. Interact.}} \bibinfo{volume}{5}, \bibinfo{number}{{CSCW1}} (\bibinfo{year}{2021}), \bibinfo{pages}{175:1--175:28}.
\newblock
\href{https://doi.org/10.1145/3449249}{doi:\nolinkurl{10.1145/3449249}}


\bibitem[Gerosa et~al\mbox{.}(2021)]%
        {gerosa2021}
\bibfield{author}{\bibinfo{person}{Marco~Aur{\'{e}}lio Gerosa}, \bibinfo{person}{Igor Wiese}, \bibinfo{person}{Bianca Trinkenreich}, \bibinfo{person}{Georg Link}, \bibinfo{person}{Gregorio Robles}, \bibinfo{person}{Christoph Treude}, \bibinfo{person}{Igor Steinmacher}, {and} \bibinfo{person}{Anita Sarma}.} \bibinfo{year}{2021}\natexlab{}.
\newblock \showarticletitle{The Shifting Sands of Motivation: Revisiting What Drives Contributors in Open Source}. In \bibinfo{booktitle}{\emph{43rd {IEEE/ACM} International Conference on Software Engineering, ({ICSE} 2021)}}. \bibinfo{publisher}{{IEEE}}, \bibinfo{pages}{1046--1058}.
\newblock
\href{https://doi.org/10.1109/ICSE43902.2021.00098}{doi:\nolinkurl{10.1109/ICSE43902.2021.00098}}


\bibitem[Gilmer et~al\mbox{.}(2023)]%
        {GilmerBSCCG23}
\bibfield{author}{\bibinfo{person}{Saskia Gilmer}, \bibinfo{person}{Avinash Bhat}, \bibinfo{person}{Shuvam Shah}, \bibinfo{person}{Kevin Cherry}, \bibinfo{person}{Jinghui Cheng}, {and} \bibinfo{person}{Jin L.~C. Guo}.} \bibinfo{year}{2023}\natexlab{}.
\newblock \showarticletitle{{SUMMIT}: {Scaffolding} {Open} {Source} {Software} {Issue} {Discussion} {Through} {Summarization}}.
\newblock \bibinfo{journal}{\emph{Proc. ACM Hum. Comput. Interact.}} \bibinfo{volume}{7}, \bibinfo{number}{CSCW2} (\bibinfo{year}{2023}), \bibinfo{pages}{1--27}.
\newblock
\href{https://doi.org/10.1145/3610088}{doi:\nolinkurl{10.1145/3610088}}


\bibitem[Hardy and Leiba-O’Sullivan(1998)]%
        {Hardy_LeibaOSullivan_1998}
\bibfield{author}{\bibinfo{person}{Cynthia Hardy} {and} \bibinfo{person}{Sharon Leiba-O’Sullivan}.} \bibinfo{year}{1998}\natexlab{}.
\newblock \showarticletitle{The {Power} {Behind} {Empowerment}: {Implications} for {Research} and {Practice}}.
\newblock \bibinfo{journal}{\emph{Human Relations}} \bibinfo{volume}{51}, \bibinfo{number}{4} (\bibinfo{year}{1998}), \bibinfo{pages}{451--483}.
\newblock
\href{https://doi.org/10.1177/001872679805100402}{doi:\nolinkurl{10.1177/001872679805100402}}


\bibitem[Hars and Ou(2002)]%
        {alexander_hars_working_2002}
\bibfield{author}{\bibinfo{person}{Alexander Hars} {and} \bibinfo{person}{Shaosong Ou}.} \bibinfo{year}{2002}\natexlab{}.
\newblock \showarticletitle{Working for Free? Motivations for Participating in Open-Source Projects}.
\newblock \bibinfo{journal}{\emph{Int. J. Electron. Commer.}} \bibinfo{volume}{6}, \bibinfo{number}{3} (\bibinfo{year}{2002}), \bibinfo{pages}{25--39}.
\newblock
\href{https://doi.org/10.1080/10864415.2002.11044241}{doi:\nolinkurl{10.1080/10864415.2002.11044241}}


\bibitem[Haskel and Graham(2016)]%
        {10.1145/2948076.2948090}
\bibfield{author}{\bibinfo{person}{Lisa~Frances Haskel} {and} \bibinfo{person}{Paula Graham}.} \bibinfo{year}{2016}\natexlab{}.
\newblock \showarticletitle{Whats {GNU} got to do with it?: participatory design, infrastructuring and free/open source software}. In \bibinfo{booktitle}{\emph{Proceedings of the 14th Participatory Design Conference: Short Papers, Interactive Exhibitions, Workshops - Volume 2}}. \bibinfo{publisher}{ACM}, \bibinfo{address}{New York, USA}, \bibinfo{pages}{17--20}.
\newblock
\href{https://doi.org/10.1145/2948076.2948090}{doi:\nolinkurl{10.1145/2948076.2948090}}


\bibitem[Haugaard(2012)]%
        {Haugaard_2012}
\bibfield{author}{\bibinfo{person}{Mark Haugaard}.} \bibinfo{year}{2012}\natexlab{}.
\newblock \showarticletitle{Rethinking the four dimensions of power: domination and empowerment}.
\newblock \bibinfo{journal}{\emph{Journal of Political Power}} \bibinfo{volume}{5}, \bibinfo{number}{1} (\bibinfo{date}{April} \bibinfo{year}{2012}), \bibinfo{pages}{33--54}.
\newblock
\showISSN{2158-379X}
\href{https://doi.org/10.1080/2158379X.2012.660810}{doi:\nolinkurl{10.1080/2158379X.2012.660810}}


\bibitem[Haugaard(2021)]%
        {haugaard_four_2021}
\bibfield{author}{\bibinfo{person}{Mark Haugaard}.} \bibinfo{year}{2021}\natexlab{}.
\newblock \showarticletitle{The four dimensions of power: conflict and democracy}.
\newblock \bibinfo{journal}{\emph{Journal of Political Power}} \bibinfo{volume}{14}, \bibinfo{number}{1} (\bibinfo{date}{Jan.} \bibinfo{year}{2021}), \bibinfo{pages}{153--175}.
\newblock
\showISSN{2158-379X}
\href{https://doi.org/10.1080/2158379X.2021.1878411}{doi:\nolinkurl{10.1080/2158379X.2021.1878411}}


\bibitem[Hayashi et~al\mbox{.}(2013)]%
        {hayashi_why_2013}
\bibfield{author}{\bibinfo{person}{Hironori Hayashi}, \bibinfo{person}{Akinori Ihara}, \bibinfo{person}{Akito Monden}, {and} \bibinfo{person}{Ken-ichi Matsumoto}.} \bibinfo{year}{2013}\natexlab{}.
\newblock \showarticletitle{Why is collaboration needed in OSS projects? a case study of eclipse project}. In \bibinfo{booktitle}{\emph{Proceedings of the 2013 International Workshop on Social Software Engineering}} \emph{(\bibinfo{series}{SSE 2013})}. \bibinfo{publisher}{ACM}, \bibinfo{address}{New York, USA}, \bibinfo{pages}{17–20}.
\newblock
\showISBNx{9781450323130}
\href{https://doi.org/10.1145/2501535.2501539}{doi:\nolinkurl{10.1145/2501535.2501539}}


\bibitem[Hellman et~al\mbox{.}(2022)]%
        {Hellman_Chen_Uddin_Cheng_Guo_2022}
\bibfield{author}{\bibinfo{person}{Jazlyn Hellman}, \bibinfo{person}{Jiahao Chen}, \bibinfo{person}{Md.~Sami Uddin}, \bibinfo{person}{Jinghui Cheng}, {and} \bibinfo{person}{Jin L.~C. Guo}.} \bibinfo{year}{2022}\natexlab{}.
\newblock \showarticletitle{Characterizing user behaviors in open-source software user forums: an empirical study}. In \bibinfo{booktitle}{\emph{Proceedings of the 15th International Conference on Cooperative and Human Aspects of Software Engineering}} \emph{(\bibinfo{series}{CHASE '22})}. \bibinfo{publisher}{ACM}, \bibinfo{address}{New York, USA}, \bibinfo{pages}{46–55}.
\newblock
\showISBNx{9781450393423}
\href{https://doi.org/10.1145/3528579.3529178}{doi:\nolinkurl{10.1145/3528579.3529178}}


\bibitem[Hellman et~al\mbox{.}(2021)]%
        {hellman_facilitating_2021}
\bibfield{author}{\bibinfo{person}{Jazlyn Hellman}, \bibinfo{person}{Jinghui Cheng}, {and} \bibinfo{person}{Jin~L.C. Guo}.} \bibinfo{year}{2021}\natexlab{}.
\newblock \showarticletitle{Facilitating Asynchronous Participatory Design of Open Source Software: Bringing End Users into the Loop}. In \bibinfo{booktitle}{\emph{Extended Abstracts of the 2021 CHI Conference on Human Factors in Computing Systems}} \emph{(\bibinfo{series}{CHI EA '21})}. \bibinfo{publisher}{ACM}, \bibinfo{address}{New York, USA}, Article \bibinfo{articleno}{438}, \bibinfo{numpages}{7}~pages.
\newblock
\showISBNx{9781450380959}
\href{https://doi.org/10.1145/3411763.3451643}{doi:\nolinkurl{10.1145/3411763.3451643}}


\bibitem[Iivari(2009)]%
        {Iivari09}
\bibfield{author}{\bibinfo{person}{Netta Iivari}.} \bibinfo{year}{2009}\natexlab{}.
\newblock \showarticletitle{"{Constructing} the users" in open source software development: {An} interpretive case study of user participation}.
\newblock \bibinfo{journal}{\emph{Inf. Technol. People}} \bibinfo{volume}{22}, \bibinfo{number}{2} (\bibinfo{year}{2009}), \bibinfo{pages}{132--156}.
\newblock
\href{https://doi.org/10.1108/09593840910962203}{doi:\nolinkurl{10.1108/09593840910962203}}


\bibitem[Iivari(2011)]%
        {iivari_participatory_2011}
\bibfield{author}{\bibinfo{person}{Netta Iivari}.} \bibinfo{year}{2011}\natexlab{}.
\newblock \showarticletitle{Participatory design in {OSS} development: interpretive case studies in company and community {OSS} development contexts}.
\newblock \bibinfo{journal}{\emph{Behav. Inf. Technol.}} \bibinfo{volume}{30}, \bibinfo{number}{3} (\bibinfo{year}{2011}), \bibinfo{pages}{309--323}.
\newblock
\href{https://doi.org/10.1080/0144929X.2010.503351}{doi:\nolinkurl{10.1080/0144929X.2010.503351}}


\bibitem[Jahn et~al\mbox{.}(2024)]%
        {JahnE0BNMW24}
\bibfield{author}{\bibinfo{person}{Leonie Jahn}, \bibinfo{person}{Philip Engelbutzeder}, \bibinfo{person}{Dave Randall}, \bibinfo{person}{Yannick Bollmann}, \bibinfo{person}{Vasilis Ntouros}, \bibinfo{person}{Lea~Katharina Michel}, {and} \bibinfo{person}{Volker Wulf}.} \bibinfo{year}{2024}\natexlab{}.
\newblock \showarticletitle{In Between Users and Developers: Serendipitous Connections and Intermediaries in Volunteer-Driven Open-Source Software Development}. In \bibinfo{booktitle}{\emph{Proceedings of the 2024 CHI Conference on Human Factors in Computing Systems}} \emph{(\bibinfo{series}{CHI '24})}. \bibinfo{publisher}{ACM}, \bibinfo{address}{New York, USA}, Article \bibinfo{articleno}{924}, \bibinfo{numpages}{15}~pages.
\newblock
\showISBNx{9798400703300}
\href{https://doi.org/10.1145/3613904.3642541}{doi:\nolinkurl{10.1145/3613904.3642541}}


\bibitem[Jamieson et~al\mbox{.}(2022)]%
        {jamieson_2022}
\bibfield{author}{\bibinfo{person}{Jack Jamieson}, \bibinfo{person}{Eureka Foong}, {and} \bibinfo{person}{Naomi Yamashita}.} \bibinfo{year}{2022}\natexlab{}.
\newblock \showarticletitle{Maintaining Values: Navigating Diverse Perspectives in Value-Charged Discussions in Open Source Development}.
\newblock \bibinfo{journal}{\emph{Proc. {ACM} Hum. Comput. Interact.}} \bibinfo{volume}{6}, \bibinfo{number}{{CSCW2}} (\bibinfo{year}{2022}), \bibinfo{pages}{1--28}.
\newblock
\href{https://doi.org/10.1145/3555550}{doi:\nolinkurl{10.1145/3555550}}


\bibitem[Jergensen et~al\mbox{.}(2011)]%
        {Jergensen2011}
\bibfield{author}{\bibinfo{person}{Corey Jergensen}, \bibinfo{person}{Anita Sarma}, {and} \bibinfo{person}{Patrick Wagstrom}.} \bibinfo{year}{2011}\natexlab{}.
\newblock \showarticletitle{The onion patch: migration in open source ecosystems}. In \bibinfo{booktitle}{\emph{Proceedings of the 19th ACM SIGSOFT Symposium and the 13th European Conference on Foundations of Software Engineering}} \emph{(\bibinfo{series}{ESEC/FSE '11})}. \bibinfo{publisher}{ACM}, \bibinfo{address}{New York, USA}, \bibinfo{pages}{70–80}.
\newblock
\showISBNx{9781450304436}
\href{https://doi.org/10.1145/2025113.2025127}{doi:\nolinkurl{10.1145/2025113.2025127}}


\bibitem[Kemp(1984)]%
        {Kemp_1984}
\bibfield{author}{\bibinfo{person}{Peter Kemp}.} \bibinfo{year}{1984}\natexlab{}.
\newblock \showarticletitle{Review of {Michel} {Foucault}. {Beyond} {Structuralism} and {Hermeneutics}}.
\newblock \bibinfo{journal}{\emph{History and Theory}} \bibinfo{volume}{23}, \bibinfo{number}{1} (\bibinfo{year}{1984}), \bibinfo{pages}{84--105}.
\newblock
\showISSN{0018-2656}
\href{https://doi.org/10.2307/2504973}{doi:\nolinkurl{10.2307/2504973}}


\bibitem[Lakhani and Wolf(2003)]%
        {Lakhani_Wolf_2003}
\bibfield{author}{\bibinfo{person}{Karim Lakhani} {and} \bibinfo{person}{Robert~G. Wolf}.} \bibinfo{year}{2003}\natexlab{}.
\newblock \showarticletitle{Why Hackers Do What They Do: Understanding Motivation and Effort in Free/Open Source Software Projects}.
\newblock \bibinfo{journal}{\emph{SSRN Electronic Journal}} (\bibinfo{year}{2003}).
\newblock
\showISSN{1556-5068}
\href{https://doi.org/10.2139/ssrn.443040}{doi:\nolinkurl{10.2139/ssrn.443040}}


\bibitem[Lobel(2001)]%
        {Lobel_2001}
\bibfield{author}{\bibinfo{person}{Orly Lobel}.} \bibinfo{year}{2001}\natexlab{}.
\newblock \showarticletitle{Agency and {Coercion} in {Labor} and {Employment} {Relations}: {Four} {Dimensions} of {Power} in {Shifting} {Patterns} of {Work}}.
\newblock \bibinfo{journal}{\emph{University of Pennsylvania Journal of Labor and Employment Law}} \bibinfo{volume}{4}, \bibinfo{number}{1} (\bibinfo{year}{2001}), \bibinfo{pages}{121--194}.
\newblock


\bibitem[Lukes(1974)]%
        {lukes1974power}
\bibfield{author}{\bibinfo{person}{Steven Lukes}.} \bibinfo{year}{1974}\natexlab{}.
\newblock \bibinfo{booktitle}{\emph{Power: {A} radical view}}.
\newblock \bibinfo{publisher}{Macmillan}, \bibinfo{address}{London, UK}.
\newblock


\bibitem[McDonald and Goggins(2013)]%
        {mcdonald_performance_2013}
\bibfield{author}{\bibinfo{person}{Nora McDonald} {and} \bibinfo{person}{Sean Goggins}.} \bibinfo{year}{2013}\natexlab{}.
\newblock \showarticletitle{Performance and participation in open source software on GitHub}. In \bibinfo{booktitle}{\emph{CHI '13 Extended Abstracts on Human Factors in Computing Systems}} \emph{(\bibinfo{series}{CHI EA '13})}. \bibinfo{publisher}{ACM}, \bibinfo{address}{New York, USA}, \bibinfo{pages}{139–144}.
\newblock
\showISBNx{9781450319522}
\href{https://doi.org/10.1145/2468356.2468382}{doi:\nolinkurl{10.1145/2468356.2468382}}


\bibitem[Nichols and Twidale(2003)]%
        {nichols_usability_2003}
\bibfield{author}{\bibinfo{person}{David~M. Nichols} {and} \bibinfo{person}{Michael~B. Twidale}.} \bibinfo{year}{2003}\natexlab{}.
\newblock \showarticletitle{The {Usability} of {Open} {Source} {Software}}.
\newblock \bibinfo{journal}{\emph{First Monday}} \bibinfo{volume}{8}, \bibinfo{number}{1} (\bibinfo{year}{2003}).
\newblock
\href{https://doi.org/10.5210/FM.V8I1.1018}{doi:\nolinkurl{10.5210/FM.V8I1.1018}}


\bibitem[Nielsen(1994)]%
        {nielsen_usability_1994}
\bibfield{author}{\bibinfo{person}{Jakob Nielsen}.} \bibinfo{year}{1994}\natexlab{}.
\newblock \bibinfo{booktitle}{\emph{Usability {Engineering}}}.
\newblock \bibinfo{publisher}{Morgan Kaufmann Publishers Inc.}, \bibinfo{address}{San Francisco, CA, USA}.
\newblock
\showISBNx{978-0-08-052029-2}


\bibitem[Nielsen(2024)]%
        {nielsen_10_2020}
\bibfield{author}{\bibinfo{person}{Jakob Nielsen}.} \bibinfo{year}{2024}\natexlab{}.
\newblock \bibinfo{title}{10 {Usability} {Heuristics} for {User} {Interface} {Design}}.
\newblock
\urldef\tempurl%
\url{https://www.nngroup.com/articles/ten-usability-heuristics/}
\showURL{%
\tempurl}


\bibitem[Programme(2023)]%
        {undp_2023}
\bibfield{author}{\bibinfo{person}{United Nations~Development Programme}.} \bibinfo{year}{2023}\natexlab{}.
\newblock \bibinfo{title}{Accelerating {The} {SDGs} {Through} {Digital} {Public} {Infrastructure}}.
\newblock
\urldef\tempurl%
\url{https://www.undp.org/publications/accelerating-sdgs-through-digital-public-infrastructure-compendium-potential-digital-public-infrastructure}
\showURL{%
\tempurl}


\bibitem[Prost et~al\mbox{.}(2015)]%
        {ProstMT15}
\bibfield{author}{\bibinfo{person}{Sebastian Prost}, \bibinfo{person}{Elke~E. Mattheiss}, {and} \bibinfo{person}{Manfred Tscheligi}.} \bibinfo{year}{2015}\natexlab{}.
\newblock \showarticletitle{From {Awareness} to {Empowerment}: {Using} {Design} {Fiction} to {Explore} {Paths} towards a {Sustainable} {Energy} {Future}}. In \bibinfo{booktitle}{\emph{Proceedings of the 18th {ACM} {Conference} on {Computer} {Supported} {Cooperative} {Work} \& {Social} {Computing}}} \emph{(\bibinfo{series}{CSCW '15})}. \bibinfo{publisher}{ACM}, \bibinfo{address}{New York, USA}, \bibinfo{pages}{1649--1658}.
\newblock
\href{https://doi.org/10.1145/2675133.2675281}{doi:\nolinkurl{10.1145/2675133.2675281}}


\bibitem[Rajanen and Iivari(2015)]%
        {rajanen_power_2015}
\bibfield{author}{\bibinfo{person}{Mikko Rajanen} {and} \bibinfo{person}{Netta Iivari}.} \bibinfo{year}{2015}\natexlab{}.
\newblock \showarticletitle{Power, Empowerment and Open Source Usability}. In \bibinfo{booktitle}{\emph{Proceedings of the 33rd Annual ACM Conference on Human Factors in Computing Systems}} \emph{(\bibinfo{series}{CHI '15})}. \bibinfo{publisher}{ACM}, \bibinfo{address}{New York, USA}, \bibinfo{pages}{3413–3422}.
\newblock
\showISBNx{9781450331456}
\href{https://doi.org/10.1145/2702123.2702441}{doi:\nolinkurl{10.1145/2702123.2702441}}


\bibitem[Rajanen et~al\mbox{.}(2012)]%
        {rajanen_introducing_2012}
\bibfield{author}{\bibinfo{person}{Mikko Rajanen}, \bibinfo{person}{Netta Iivari}, {and} \bibinfo{person}{Eino Keskitalo}.} \bibinfo{year}{2012}\natexlab{}.
\newblock \showarticletitle{Introducing usability activities into open source software development projects: a participative approach}. In \bibinfo{booktitle}{\emph{Nordic Conference on Human-Computer Interaction, (NordiCHI '12)}}. \bibinfo{publisher}{{ACM}}, \bibinfo{address}{New York, USA}, \bibinfo{pages}{683--692}.
\newblock
\href{https://doi.org/10.1145/2399016.2399120}{doi:\nolinkurl{10.1145/2399016.2399120}}


\bibitem[Raza et~al\mbox{.}(2010)]%
        {raza_improvement_2010}
\bibfield{author}{\bibinfo{person}{Arif Raza}, \bibinfo{person}{Luiz~Fernando Capretz}, {and} \bibinfo{person}{Faheem Ahmed}.} \bibinfo{year}{2010}\natexlab{}.
\newblock \showarticletitle{Improvement of Open Source Software Usability: An Empirical Evaluation from Developers' Perspective}.
\newblock \bibinfo{journal}{\emph{Adv. Softw. Eng.}}  \bibinfo{volume}{2010} (\bibinfo{year}{2010}), \bibinfo{pages}{517532:1--517532:12}.
\newblock
\href{https://doi.org/10.1155/2010/517532}{doi:\nolinkurl{10.1155/2010/517532}}


\bibitem[Rosner et~al\mbox{.}(2016)]%
        {Rosner2016}
\bibfield{author}{\bibinfo{person}{Daniela~K. Rosner}, \bibinfo{person}{Saba Kawas}, \bibinfo{person}{Wenqi Li}, \bibinfo{person}{Nicole Tilly}, {and} \bibinfo{person}{Yi{-}Chen Sung}.} \bibinfo{year}{2016}\natexlab{}.
\newblock \showarticletitle{Out of Time, Out of Place: Reflections on Design Workshops as a Research Method}. In \bibinfo{booktitle}{\emph{Proceedings of the 19th {ACM} Conference on Computer-Supported Cooperative Work {\&} Social Computing, ({CSCW} 2016)}}. \bibinfo{publisher}{{ACM}}, \bibinfo{address}{New York, USA}, \bibinfo{pages}{1129--1139}.
\newblock
\href{https://doi.org/10.1145/2818048.2820021}{doi:\nolinkurl{10.1145/2818048.2820021}}


\bibitem[Sanei and Cheng(2024)]%
        {SaneiC24}
\bibfield{author}{\bibinfo{person}{Arghavan Sanei} {and} \bibinfo{person}{Jinghui Cheng}.} \bibinfo{year}{2024}\natexlab{}.
\newblock \showarticletitle{Characterizing {Usability} {Issue} {Discussions} in {Open} {Source} {Software} {Projects}}.
\newblock \bibinfo{journal}{\emph{Proc. ACM Hum. Comput. Interact.}} \bibinfo{volume}{8}, \bibinfo{number}{CSCW1} (\bibinfo{year}{2024}), \bibinfo{pages}{1--26}.
\newblock
\href{https://doi.org/10.1145/3637307}{doi:\nolinkurl{10.1145/3637307}}


\bibitem[Santos et~al\mbox{.}(2022)]%
        {santos2022}
\bibfield{author}{\bibinfo{person}{Fabio Santos}, \bibinfo{person}{Bianca Trinkenreich}, \bibinfo{person}{Jo\~{a}o~Felipe Pimentel}, \bibinfo{person}{Igor Wiese}, \bibinfo{person}{Igor Steinmacher}, \bibinfo{person}{Anita Sarma}, {and} \bibinfo{person}{Marco~A. Gerosa}.} \bibinfo{year}{2022}\natexlab{}.
\newblock \showarticletitle{How to Choose a Task? Mismatches in Perspectives of Newcomers and Existing Contributors}. In \bibinfo{booktitle}{\emph{Proceedings of the 16th ACM / IEEE International Symposium on Empirical Software Engineering and Measurement}} \emph{(\bibinfo{series}{ESEM '22})}. \bibinfo{publisher}{ACM}, \bibinfo{address}{New York, USA}, \bibinfo{pages}{114–124}.
\newblock
\showISBNx{9781450394277}
\href{https://doi.org/10.1145/3544902.3546236}{doi:\nolinkurl{10.1145/3544902.3546236}}


\bibitem[Smith et~al\mbox{.}(2020)]%
        {SmithYSHTZ20}
\bibfield{author}{\bibinfo{person}{C.~Estelle Smith}, \bibinfo{person}{Bowen Yu}, \bibinfo{person}{Anjali Srivastava}, \bibinfo{person}{Aaron Halfaker}, \bibinfo{person}{Loren Terveen}, {and} \bibinfo{person}{Haiyi Zhu}.} \bibinfo{year}{2020}\natexlab{}.
\newblock \showarticletitle{Keeping Community in the Loop: Understanding Wikipedia Stakeholder Values for Machine Learning-Based Systems}. In \bibinfo{booktitle}{\emph{Proceedings of the 2020 CHI Conference on Human Factors in Computing Systems}} \emph{(\bibinfo{series}{CHI '20})}. \bibinfo{publisher}{ACM}, \bibinfo{address}{New York, USA}, \bibinfo{pages}{1–14}.
\newblock
\showISBNx{9781450367080}
\href{https://doi.org/10.1145/3313831.3376783}{doi:\nolinkurl{10.1145/3313831.3376783}}


\bibitem[Steinmacher et~al\mbox{.}(2015)]%
        {SteinmacherCGR15}
\bibfield{author}{\bibinfo{person}{Igor Steinmacher}, \bibinfo{person}{Tayana Conte}, \bibinfo{person}{Marco~Aur\'{e}lio Gerosa}, {and} \bibinfo{person}{David Redmiles}.} \bibinfo{year}{2015}\natexlab{}.
\newblock \showarticletitle{Social Barriers Faced by Newcomers Placing Their First Contribution in Open Source Software Projects}. In \bibinfo{booktitle}{\emph{Proceedings of the 18th ACM Conference on Computer Supported Cooperative Work \& Social Computing}} \emph{(\bibinfo{series}{CSCW '15})}. \bibinfo{publisher}{ACM}, \bibinfo{address}{New York, USA}, \bibinfo{pages}{1379–1392}.
\newblock
\showISBNx{9781450329224}
\href{https://doi.org/10.1145/2675133.2675215}{doi:\nolinkurl{10.1145/2675133.2675215}}


\bibitem[Terry et~al\mbox{.}(2010)]%
        {terry_perceptions_2010}
\bibfield{author}{\bibinfo{person}{Michael~A. Terry}, \bibinfo{person}{Matthew Kay}, {and} \bibinfo{person}{Ben Lafreniere}.} \bibinfo{year}{2010}\natexlab{}.
\newblock \showarticletitle{Perceptions and practices of usability in the free/open source software ({FoSS}) community}. In \bibinfo{booktitle}{\emph{Proceedings of the 28th {International} {Conference} on {Human} {Factors} in {Computing} {Systems}}} \emph{(\bibinfo{series}{{CHI} '10})}. \bibinfo{publisher}{ACM}, \bibinfo{address}{New York, USA}, \bibinfo{pages}{999--1008}.
\newblock
\href{https://doi.org/10.1145/1753326.1753476}{doi:\nolinkurl{10.1145/1753326.1753476}}


\bibitem[Tsay et~al\mbox{.}(2014)]%
        {tsay_lets_2014}
\bibfield{author}{\bibinfo{person}{Jason Tsay}, \bibinfo{person}{Laura Dabbish}, {and} \bibinfo{person}{James Herbsleb}.} \bibinfo{year}{2014}\natexlab{}.
\newblock \showarticletitle{Let's talk about it: evaluating contributions through discussion in GitHub}. In \bibinfo{booktitle}{\emph{Proceedings of the 22nd ACM SIGSOFT International Symposium on Foundations of Software Engineering}} \emph{(\bibinfo{series}{FSE 2014})}. \bibinfo{publisher}{ACM}, \bibinfo{address}{New York, USA}, \bibinfo{pages}{144–154}.
\newblock
\showISBNx{9781450330565}
\href{https://doi.org/10.1145/2635868.2635882}{doi:\nolinkurl{10.1145/2635868.2635882}}


\bibitem[Vaismoradi et~al\mbox{.}(2013)]%
        {Vaismoradi2013}
\bibfield{author}{\bibinfo{person}{Mojtaba Vaismoradi}, \bibinfo{person}{Hannele Turunen}, {and} \bibinfo{person}{Terese Bondas}.} \bibinfo{year}{2013}\natexlab{}.
\newblock \showarticletitle{Content analysis and thematic analysis: {Implications} for conducting a qualitative descriptive study}.
\newblock \bibinfo{journal}{\emph{Nursing \& Health Sciences}} \bibinfo{volume}{15}, \bibinfo{number}{3} (\bibinfo{year}{2013}), \bibinfo{pages}{398--405}.
\newblock
\href{https://doi.org/10.1111/nhs.12048}{doi:\nolinkurl{10.1111/nhs.12048}}


\bibitem[von Krogh et~al\mbox{.}(2012)]%
        {von_Krogh_Haefliger_Spaeth_Wallin_2012}
\bibfield{author}{\bibinfo{person}{Georg von Krogh}, \bibinfo{person}{Stefan Haefliger}, \bibinfo{person}{Sebastian Spaeth}, {and} \bibinfo{person}{Martin~W. Wallin}.} \bibinfo{year}{2012}\natexlab{}.
\newblock \showarticletitle{Carrots and Rainbows: Motivation and Social Practice in Open Source Software Development}.
\newblock \bibinfo{journal}{\emph{MIS Quarterly}} \bibinfo{volume}{36}, \bibinfo{number}{2} (\bibinfo{year}{2012}), \bibinfo{pages}{649--676}.
\newblock
\showISSN{02767783}
\href{https://doi.org/10.2307/41703471}{doi:\nolinkurl{10.2307/41703471}}


\bibitem[Vyas et~al\mbox{.}(2023)]%
        {VyasKC23}
\bibfield{author}{\bibinfo{person}{Dhaval Vyas}, \bibinfo{person}{Awais~Hameed Khan}, {and} \bibinfo{person}{Anabelle Cooper}.} \bibinfo{year}{2023}\natexlab{}.
\newblock \showarticletitle{Democratizing Making: Scaffolding Participation Using e-Waste to Engage Under-resourced Communities in Technology Design}. In \bibinfo{booktitle}{\emph{Proceedings of the 2023 CHI Conference on Human Factors in Computing Systems}} \emph{(\bibinfo{series}{CHI '23})}. \bibinfo{publisher}{ACM}, \bibinfo{address}{New York, USA}, Article \bibinfo{articleno}{301}, \bibinfo{numpages}{16}~pages.
\newblock
\showISBNx{9781450394215}
\href{https://doi.org/10.1145/3544548.3580759}{doi:\nolinkurl{10.1145/3544548.3580759}}


\bibitem[Wang et~al\mbox{.}(2020)]%
        {wang_argulens_2020}
\bibfield{author}{\bibinfo{person}{Wenting Wang}, \bibinfo{person}{Deeksha Arya}, \bibinfo{person}{Nicole Novielli}, \bibinfo{person}{Jinghui Cheng}, {and} \bibinfo{person}{Jin~L.C. Guo}.} \bibinfo{year}{2020}\natexlab{}.
\newblock \showarticletitle{ArguLens: Anatomy of Community Opinions On Usability Issues Using Argumentation Models}. In \bibinfo{booktitle}{\emph{Proceedings of the 2020 CHI Conference on Human Factors in Computing Systems}} \emph{(\bibinfo{series}{CHI '20})}. \bibinfo{publisher}{ACM}, \bibinfo{address}{New York, USA}, \bibinfo{pages}{1–14}.
\newblock
\showISBNx{9781450367080}
\href{https://doi.org/10.1145/3313831.3376218}{doi:\nolinkurl{10.1145/3313831.3376218}}


\bibitem[Wang et~al\mbox{.}(2022)]%
        {wang_how_2020}
\bibfield{author}{\bibinfo{person}{Wenting Wang}, \bibinfo{person}{Jinghui Cheng}, {and} \bibinfo{person}{Jin L.~C. Guo}.} \bibinfo{year}{2022}\natexlab{}.
\newblock \showarticletitle{How {Do} {Open} {Source} {Software} {Contributors} {Perceive} and {Address} {Usability}?: {Valued} {Factors}, {Practices}, and {Challenges}}.
\newblock \bibinfo{journal}{\emph{{IEEE} Softw.}} \bibinfo{volume}{39}, \bibinfo{number}{1} (\bibinfo{year}{2022}), \bibinfo{pages}{76--83}.
\newblock
\href{https://doi.org/10.1109/MS.2020.3009514}{doi:\nolinkurl{10.1109/MS.2020.3009514}}


\bibitem[Wubishet et~al\mbox{.}(2013)]%
        {wubishet2013participation}
\bibfield{author}{\bibinfo{person}{Zegaye~S. Wubishet}, \bibinfo{person}{Bendik Bygstad}, {and} \bibinfo{person}{Prodromos Tsiavos}.} \bibinfo{year}{2013}\natexlab{}.
\newblock \showarticletitle{A participation paradox: Seeking the missing link between free/open source software and participatory design}.
\newblock \bibinfo{journal}{\emph{Journal of Advances in Information Technology}} \bibinfo{volume}{4}, \bibinfo{number}{4} (\bibinfo{year}{2013}), \bibinfo{pages}{181--193}.
\newblock


\bibitem[Xuan and Filkov(2014)]%
        {xuan_building_2014}
\bibfield{author}{\bibinfo{person}{Qi Xuan} {and} \bibinfo{person}{Vladimir Filkov}.} \bibinfo{year}{2014}\natexlab{}.
\newblock \showarticletitle{Building it together: synchronous development in {OSS}}. In \bibinfo{booktitle}{\emph{36th International Conference on Software Engineering, ({ICSE} '14)}}. \bibinfo{publisher}{{ACM}}, \bibinfo{address}{New York, USA}, \bibinfo{pages}{222--233}.
\newblock
\href{https://doi.org/10.1145/2568225.2568238}{doi:\nolinkurl{10.1145/2568225.2568238}}


\end{thebibliography}

\newpage
\appendix
\section{Material Used in the Design Workshops}
\label{appx:material}
\subsection{Three Personas}
\label{appx:personas}

\begin{figure}[ht!]
    \centering
    \includegraphics[width=\textwidth]{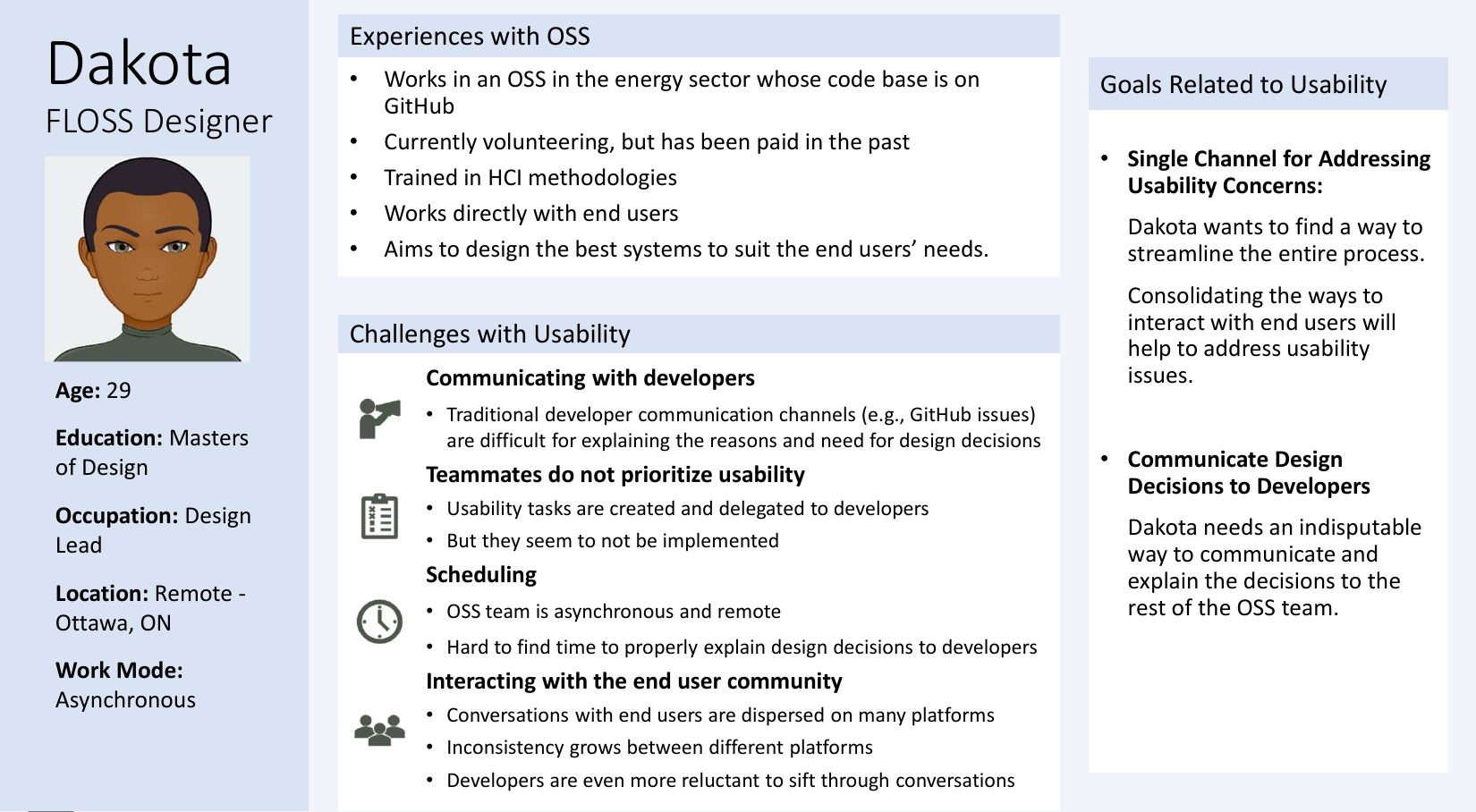}
    \caption{The designer persona, Dakota}
    \Description{}
    \label{fig:persona_dakota}
\end{figure}

\begin{figure}[ht!]
    \centering
    \includegraphics[width=\textwidth]{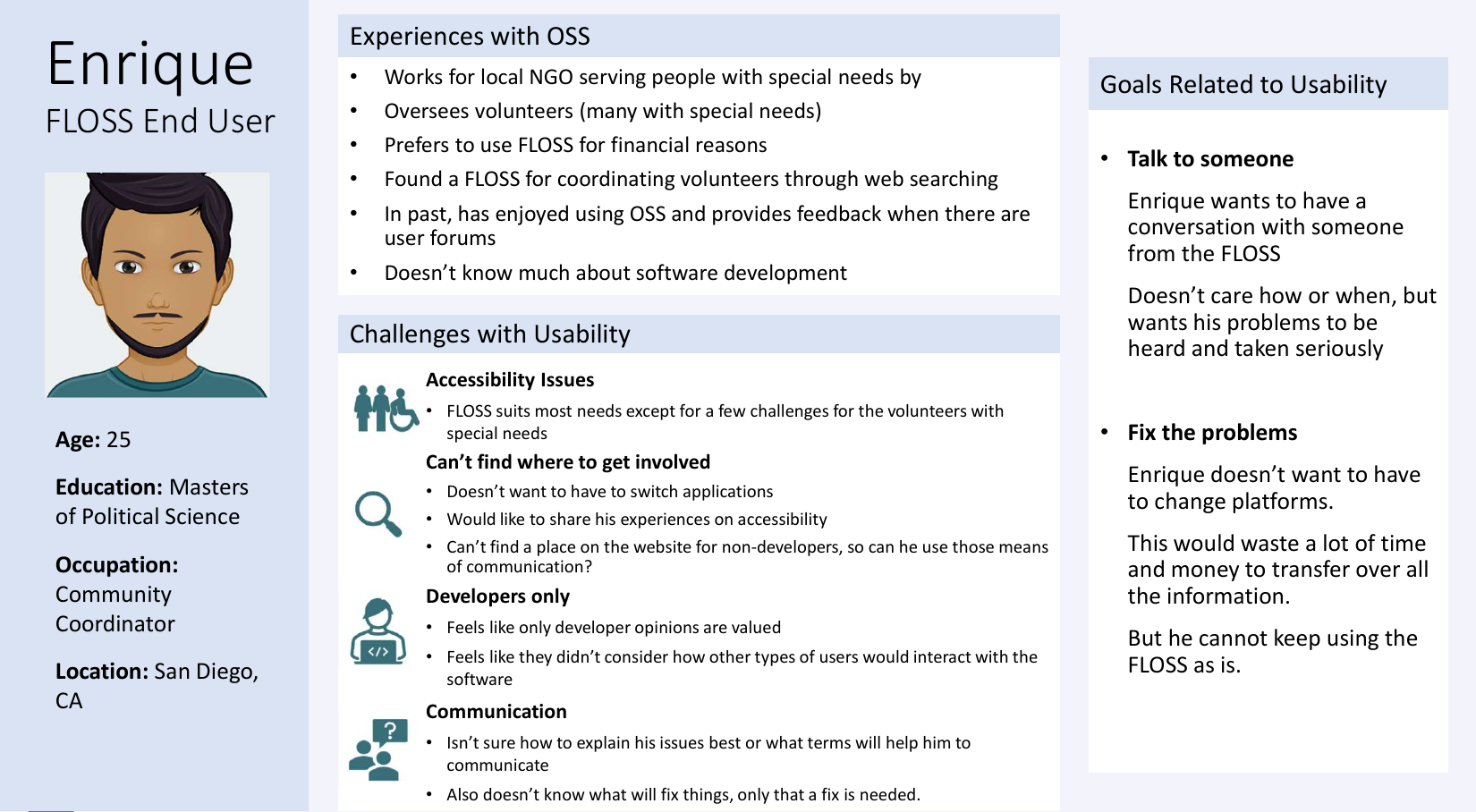}
    \caption{The end-user persona, Enrique}
    \Description{}
    \label{fig:persona_enrique}
\end{figure}

\begin{figure}[ht!]
    \centering
    \includegraphics[width=\textwidth]{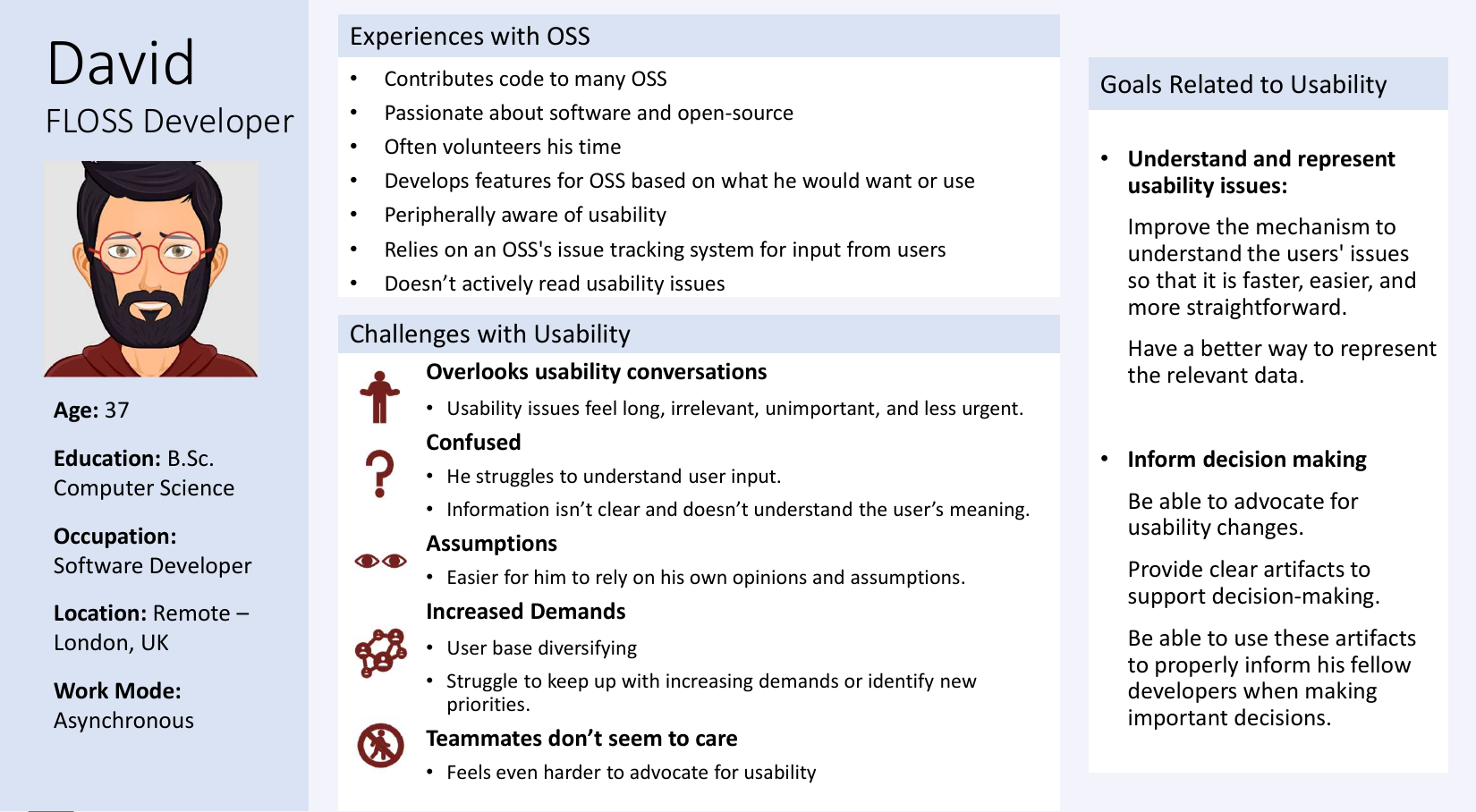}
    \caption{The developer persona, David}
    \Description{}
    \label{fig:persona_david}
\end{figure}

\subsection{Design Objective Instruction}
\label{appx:design_obj}

You and a small group of peers have been asked to work together and design a new tool to help support the inclusion of developers, designers, and end-users of free, libre, and open-source software (FLOSS) in addressing usability issues of that type of software.

Your objective is to work together to (a) identify a specific problem facing the inclusion of different stakeholders to achieve usable FLOSS and (b) design a concept to resolve the problem. As a team, create a written design pitch that illustrates the design concept and explains any relevant details and decisions. 

Some elements for you to consider including in your design pitch (you do not have to follow this
strictly) are:

\begin{itemize}
    \item The problem statement
    \item Key features
    \item Why decisions were made
    \item How the tool will benefit the users
    \item Visual and external materials (i.e., sketches, mockups, prototypes, important links, etc.)
\end{itemize}

\end{document}